\documentclass[aip,cha,numerical,groupedaddress,reprint,amsmath,amssymb]{revtex4-1}

\usepackage{graphicx}

\begin{document}

\title[Intermittency and $1/f$ noise]{Intermittency in relation with $1/f$ noise and stochastic differential equations}

\author{J.~Ruseckas}
\email{julius.ruseckas@tfai.vu.lt}
\homepage{http://www.itpa.lt/~ruseckas}
\author{B.~Kaulakys}
\affiliation{Institute of Theoretical Physics and Astronomy, Vilnius University,
A.~Go\v{s}tauto 12, LT-01108 Vilnius, Lithuania}

\begin{abstract}
One of the models of intermittency is on-off intermittency, arising due to
time-dependent forcing of a bifurcation parameter through a bifurcation point.
For on-off intermittency the power spectral density of the time-dependent
deviation from the invariant subspace in a low frequency region exhibits
$1/\sqrt{f}$ power-law noise. Here we investigate a mechanism of intermittency,
similar to the on-off intermittency, occurring in nonlinear dynamical systems
with invariant subspace. In contrast to the on-off intermittency, we consider
the case where the transverse Lyapunov exponent is zero. We show that for such
nonlinear dynamical systems the power spectral density of the deviation from
the invariant subspace can have $1/f^\beta$ form in a wide range of
frequencies. That is, such nonlinear systems exhibit $1/f$ noise. The
connection with the stochastic differential equations generating $1/f^\beta$
noise is established and analyzed, as well.
\end{abstract}

\maketitle

\begin{quotation} 
The phrase ``$1/f$ noise'' refers to the well-known empirical fact that in many
systems at low frequencies the noise spectrum exhibits an approximately $1/f$
shape. Generating mechanisms leading to $1/f^{\beta}$ noise are still an open
question. Here we analyze nonlinear dynamical systems with invariant subspace
having the transverse Lyapunov exponent equal to zero. In particular, we explore
nonlinear maps having power-law dependence on the deviation from the invariant
subspace. We demonstrate that such maps can generate signals exhibiting
$1/f^{\beta}$ noise and intermittent behavior. In contrast to known mechanism of
$1/f$ noise involving Pomeau-Manneville type maps, coefficients in the maps we
consider are not static, similarly as in the maps describing on-off
intermittency. We relate the nonlinear dynamics described by proposed maps to
$1/f$ noise models based on the nonlinear stochastic differential equations.
\end{quotation}

\section{Introduction}

Intermittency is an apparently random alternation of a signal between a
quiescent state and bursts of activity. In 1949, Batchelor and Townsend used
the word intermittency to describe their observations of the patchiness of the
fluctuating velocity field in a fully turbulent fluid.\cite{Batchelor1949}
Many natural systems display intermittent behavior, for example, turbulent
bursts in otherwise laminar fluid flows, sunspot activity, and reversals of the
geomagnetic field. Well known models of intermittency include the three types
introduced by Pomeau and Manneville,\cite{Pommeau1980} as well as
crisis-induced intermittency.\cite{Grebogi1987} A different variety of
intermittency was first reported when synchronized chaos in a coupled chaotic
oscillator system undergoes the instability as the coupling constant is changed
\cite{Fujisaka1985,Fujisaka1986}. This intermittency is now known as on-off
intermittency.
\cite{Platt1993,Heagy1994,Yamada1989,*Ott1994,*Lai1995,*Cenys1996,*Venkataramani1996,*Lai1996,*Lai1996a,*Fujisaka1998,*Harada1999,*Becker1999}
On-off intermittency appears in nonlinear dynamical systems with invariant
subspaces, where the dynamics restricted to the invariant subspace is chaotic
and the system is close to a threshold of transverse stability of the subspace.
The main difference of on-off intermittency from other types is in the
mechanism of the origin: on-off intermittency relies on the time-dependent
forcing of a bifurcation parameter through a bifurcation point; in
Pomeau-Manneville intermittency and crisis-induced intermittency the parameters
are static. 

It is known that the on-off intermittency exhibits characteristic statistics:
\cite{Yamada1986,*Yamada1990,*Fujisaka1987,*Fujisaka1993,*Suetani1999,*Fujisaka2000,*Fujisaka1997,*Miyazaki2000}
(i) the probability density function (PDF) of the magnitude of deviation $\rho$
from the invariant subspace obeys the asymptotic power-law, $\rho^{-1+\eta}$, with a
small positive exponent $\eta$, (ii) the power spectral density (PSD) of the time
series $\{\rho(t)\}$ in a low-frequency region exhibits a power-law
$1/\sqrt{f}$ dependence, and (iii) given an appropriately small threshold
$\rho_{\mathrm{th}}$, the PDF of the laminar duration $\tau$ takes an
asymptotic form $\tau^{-3/2}$ in a certain wide range of $\tau$.
\cite{Platt1993,Heagy1994}
Since on-off intermittency generates signals having $f^{-\beta}$ PSD with
$\beta=1/2$, the question arises whether a mechanism of intermittency, similar
to the on-off intermittency, can yield signals having other values of the exponent
$\beta$ of PSD, in particular $\beta=1$. The purpose of this paper is to
investigate this question.

Signals having the PSD at low frequencies $f$ of the form $S(f)\sim1/f^{\beta}$ with $\beta$ close to $1$
are commonly referred to as ``$1/f$ noise'', ``$1/f$ fluctuations'', or ``flicker noise.''
Power-law distributions of spectra
of signals with $0.5<\beta<1.5$, as well as scaling behavior in general, are
ubiquitous in physics and in many other fields, including natural phenomena,
human activities, traffics in computer networks and financial markets.
\cite{Scholarpedia2007,Weissman1988,*Barabasi1999,*Gisiger2001,*Wong2003,*Wagenmakers2004,*Newman05,*Szabo2007,*Castellano2009,*Eliazar2009,*Eliazar2010,*Perc2010,*Werner2010,*Orden2010,*Kendal2011,*Torabi2011,*Diniz2011}
Many models and theories of $1/f$ noise are not universal because
of the assumptions specific to the problem under consideration. 
Recently, the nonlinear stochastic differential equations (SDEs) generating signals
with $1/f$ noise were obtained in Refs.~\onlinecite{Kaulakys2004,*Kaulakys2006}
(see also recent papers \cite{Kaulakys2009,Ruseckas10}), starting from the point
process model of $1/f$ noise.
\cite{Kaulakys1998-1,*Kaulakys1999-1,*Kaulakys2000-1,*Kaulakys2000-2,*Gontis2004,*Kaulakys2005} 
Yet another model of $1/f$ noise involves a class of maps generating
intermittent signals. It is possible to generate power-laws and $1/f$-noise from
simple iterative maps by fine-tuning the parameters of the system at the edge of
chaos \cite{Procaccia1983,Schuster1988} where the sensitivity to initial conditions of the
logistic map is a lot milder than in the chaotic regime: the Lyapunov exponent
is zero and the sensitivity to changes in initial conditions follows a power-law.
\cite{Costa1997} Manneville \cite{Manneville1980a} first showed that, tuned
exactly, an iterative function can produce interesting behavior, power-laws and
$1/f$ PSD. This mechanism for $1/f$ noise only works for type-II and type-III
Pomeau-Manneville intermittency. \cite{Ben-Mizrachi1985} Intermittency as a
mechanism of $1/f$ noise continues to attract attention.
\cite{Laurson2006,*Pando2007,*Shinkai2012}

In this paper we consider a mechanism of intermittency,
similar to the on-off intermittency, occurring in nonlinear dynamical systems
with invariant subspace. In contrast to the on-off intermittency, we consider
the case where the transverse Lyapunov exponent is zero.
In recent years there is a growing interest in dynamical systems which are
characterized by zero Lyapunov exponents, namely, which trajectories diverge
nonexponentially. \cite{Zaslavsky2007} Chaos in such dynamical systems is
called weak chaos.
By relating nonlinear dynamics  with the $1/f$ 
noise model based on the nonlinear SDEs we show that for such
nonlinear dynamical systems the power spectral density of the deviation from
the invariant subspace can have $1/f^\beta$ form, i.e., $1/f$ noise in a wide range of frequencies. 
Thus a generalization of 
the on-off intermittency yields a new mechanism of $1/f^{\beta}$ noise with 
the asymptotically power-law PDF, originated from the commonly known phenomenon 
of the intermittency and weak chaos. 

This paper is organized as follows: in Sec.~\ref{sec:model} we propose a model
of intermittency with zero transverse Lyapunov exponent and in
Sec.~\ref{sec:examples} we present some examples of nonlinear maps exhibiting
$1/f$ noise. To obtain analytical expressions of the PDF and PSD of the
deviation from the invariant subspace, in Sec.~\ref{sec:map-SDE} we approximate
discrete maps with SDEs. Section \ref{sec:concl} summarizes our findings. 

\section{\label{sec:model}Model of intermittency with zero transverse Lyapunov exponent}

We consider two-dimensional maps having a skew product
structure: \cite{Platt1993}
\begin{equation}
x_{n+1}=F(x_{n})\,,\qquad y_{n+1}=G(x_{n},y_{n})\,.
\label{eq:map0}
\end{equation}
The function $G$ has the property $G(x,0)=0$ and, thus, $y=0$ is the invariant
subspace, while $y$ is the deviation form the invariant subspace. We assume that
the dynamics $x_{n+1}=F(x_{n})$ in~(\ref{eq:map0}) restricted  to the invariant
subspace is chaotic. If the transverse Lyapunov exponent
\begin{equation}
\lambda_{\bot}=\lim_{N\rightarrow\infty}\frac{1}{N}
\sum_{n=0}^{N-1}\ln\left|\frac{\partial G(x_{n},0)}{\partial y}\right|
\end{equation}
along an orbit on the invariant subspace converges and is less than zero, then
the invariant subspace is transversely stable with respect to this orbit. 

In this article we consider the case when $\partial G(x,0)/\partial y=1$ and,
consequently, the transverse Lyapunov exponent is zero. Furthermore, we will
assume that the two terms with the lowest powers in the expansion of the
function $G(x,y)$ in the power series of $y$ have the form 
\begin{equation}
G(x,y)=y+g(x)y^{\eta}
\label{eq:G-expansion}
\end{equation}
with $\eta>1$. This form satisfies the condition $\partial G(x,0)/\partial y=1$.
Particularly $\eta=2$, however, generally $\eta$ may be fractional, as well. 

We will consider the case where the function $g(x)$ in
Eq.~(\ref{eq:G-expansion}) is not constant and can acquire both positive and
negative values. Thus the expansion (\ref{eq:G-expansion}) leads to the the map
for small values of $y_n$
\begin{equation}
y_{n+1}=y_{n}+z_{n}y_{n}^{\eta}\,,\qquad\eta>1\,,
\label{eq:map1}
\end{equation}
where $z_{n}\equiv g(x_{n})$. It should be noted that when $\eta=1$, the map
(\ref{eq:map1}) becomes a multiplicative map $y_{n+1}=y_{n}(1+z_{n})$, which is
essentially the same as the map considered in Ref.~\onlinecite{Heagy1994} for
modeling of on-off intermittency. The map (\ref{eq:map1}) is similar to
Pomeau-Manneville map 
\begin{equation}
y_{n+1}=y_{n}+ay_{n}^{\eta}\pmod1
\label{eq:P-M}
\end{equation}
on the unit interval with one marginally unstable fixed point located at $y=0$.
\cite{Pommeau1980} The main difference from the map (\ref{eq:map1}) is that in
the Pomeau-Manneville map (\ref{eq:P-M}) the coefficient in the second term is
static. 

Let us consider the situation when $y_n>0$. If $z_{n}<0$ then the map
(\ref{eq:map1}) leads to the decrease of the deviation from the invariant
subspace $y=0$, whereas for $z_{n}>0$ the deviation $y$ grows. In contrast to
systems with nonzero transverse Lyapunov exponent, the growth or decrease of the
deviation is not exponential. In fact, if the second term on the right-hand side
of Eq.~(\ref{eq:map1}) is much smaller than the first and, consequently,
Eq.~(\ref{eq:map1}) can be approximately replaced by the differential equation
$dy/dt=y^{\eta}z$, the growth or decrease of the deviation $y$ can be described
by a $q$-exponential function with $q=\eta$. The $q$-exponential function, used
in the framework of nonextensive statistical mechanics,
\cite{Tsallis1988,Tsallis2009,Tsallis2009a} is defined as
\begin{equation}
\exp_{q}(x)\equiv[1+(1-q)x]_{+}^{\frac{1}{1-q}}\,,
\label{eq:q-exp}
\end{equation}
where $[x]_{+}\equiv\max\{x,0\}$. Thus, although the Lyapunov exponent is zero,
the map can be characterized by a nonzero $q$-generalized Lyapunov
coefficient.\cite{Costa1997,Tsallis2009a}

If the average of the variable $z$ is positive, $\langle z\rangle>0$, and there
is a global mechanism of reinjection, the map (\ref{eq:map1}) leads to the
intermittent behavior. As in on-off intermittency, the intermittent behavior
appears due to the time-dependent forcing of a bifurcation parameter through a
bifurcation point $z=0$, thus the behavior described by map (\ref{eq:map1}) can
be considered as a kind of on-off intermittency. However, on-off intermittency
is usually investigated in dynamical systems with nonzero transverse Lyapunov
exponent.

For small durations of the laminar phase, one can approximate the map
(\ref{eq:map1}) replacing $y_{n}$ in the second term on the right hand side
with initial value $y_{0}$. In this case Eq.~(\ref{eq:map1}) describes a random
walk with drift. Since the average displacement due to the diffusion grows as
$\sqrt{t}$ and the displacement due to drift term is proportional to $t$, for
small enough durations $t$ the diffusion is more important than the drift. It
is known that for the unbiased random walk the distribution of the first return
times has the power-law exponent $-3/2$. \cite{Redner2001} Therefore, for small
enough durations $t$ one can expect to observe the power-law form, $t^{-3/2}$,
of the PDF of the laminar phase durations, the same as in on-off intermittency. 

The first two terms in the expansion (\ref{eq:G-expansion}) do not allow to
determine uniquely the PDF of the deviation $y$. In order to determine PDF of
$y$ and PSD of the series $\{y_{n}\}$, we need to take into account more terms
in the expansion of the function $G(x,y)$ in the power series of $y$. One of
the possibilities that we will consider is for the third term in the expansion
to be equal to $\gamma y^{2\eta-1}$ (note, that $2\eta-1>\eta$ when $\eta>1$),
leading to the map
\begin{equation}
y_{n+1}=y_{n}+z_{n}y_{n}^{\eta}+\gamma y_{n}^{2\eta-1}\,.
\label{eq:map2}
\end{equation}
Particularly, for $\eta=2$, $2\eta-1=3$ and Eqs. (\ref{eq:G-expansion}) and
(\ref{eq:map2}) display  simple Taylor expansions.
Note, that a mechanism of reinjection operates at large values of $y$ and does
not change Eq.~(\ref{eq:map2}), written for small values of $y$ close to the 
inveriant subspace.

\subsection{$q$-exponential transformation of random walk}

Another example of the function $G(x,y)$ having the expansion in
the power series of $y$ as in Eq.~(\ref{eq:map1}) can be obtained
according to the following consideration: In Ref.~\onlinecite{Heagy1994}
a map of the form
\begin{equation}
y_{n+1}=w_{n}y_{n}
\label{eq:map-on-off}
\end{equation}
was considered as a model of on-off intermittency. In the log domain
this map transforms to
\begin{equation}
s_{n+1}=s_{n}+z_{n}\,,
\label{eq:map-on-off-log}
\end{equation}
where $s_{n}=\ln y_{n}$ and $z_{n}=\ln w_{n}$. The critical condition for the
onset of on-off intermittency is the condition for unbiased random walk,
$\langle z\rangle=0$. One of the reasons for intermittent behavior is highly
non-linear relation $y_{n}=e^{s_{n}}$ between $s_{n}$ and $y_{n}$. We can
expect intermittent behavior also using other nonlinear functions instead of
the exponential function. One of the generalizations of the exponential
function, which corresponds to the differential equation $dy/ds=y$, is the
$q$-exponential function (\ref{eq:q-exp}) obeying the equation
$dy/ds=y^{\eta}$. Thus, instead of $y_{n}=e^{s_{n}}$ we will consider a
relation $y_{n}=\exp_{\eta}(s_{n})$, leading to a map of the form
\begin{equation}
y_{n+1}=\exp_{\eta}(\ln_{\eta}(y_{n})+z_{n})=
(y_{n}^{1-\eta}+(1-\eta)z_{n})^{\frac{1}{1-\eta}}\,,
\label{eq:map-q-exp}
\end{equation}
where the $q$-logarithm, defined as \cite{Tsallis2009} 
\begin{equation}
\ln_{q}(x)=\frac{x^{1-q}-1}{1-q}, 
\end{equation}
is a function inverse to $q$-exponential function. Expanding the map
(\ref{eq:map-q-exp}) in power series of $y$ we get Eq.~(\ref{eq:map1}). 

The $q$-exponential function $\exp_{\eta}(s_{n})$ tends to infinity as $s_{n}$
approaches $1/(\eta-1)$ and the variable $y_n=\exp_{\eta}(s_{n})$ can be
introduced only when $s_n$ does not reach $1/(\eta-1)$. This can be achieved by
modifying the map (\ref{eq:map-on-off-log}) for the values of $s_{n}$ close to
$1/(\eta-1)$ in order to avoid reaching this value. The modification of the map
(\ref{eq:map-on-off-log}) changes also the map (\ref{eq:map-q-exp}) for large
values of $y_{n}$, not allowing for the value of the expression
$y_{n}^{1-\eta}+(1-\eta)z_{n}$ to become zero.

\section{\label{sec:examples}Numerical examples}

In this Section we present some examples of the map
(\ref{eq:map0}) with the function $G(x,y)$ whose behavior for small values of
$y$ is described by Eq.~(\ref{eq:map2}) or Eq.~(\ref{eq:map-q-exp}). Let us
consider the map (\ref{eq:map2}) with $\eta=2$, $\gamma=0.5$ when the variable
$z_{n}$ has the average $\langle z\rangle=5\times10^{-5}$ and the variance
$\langle(z-\langle z\rangle)^{2}\rangle=1$. The parameters of the map are chosen
taking into account equations from Sec.~\ref{sec:map-SDE}. The chosen value of
the average $\langle z\rangle$ is close to the critical value for the onset of
intermittency $\langle z\rangle = 0$ and is much smaller than the standard
deviation of the variable $z_n$. As a mechanism of reinjection we use a
reflection at $y=0.5$, leading to the map
\begin{equation}
y_{n+1}=0.5-|y_{n}+z_{n}y_{n}^{2}+0.5y_{n}^{3}-0.5|\,.
\label{eq:ex-map-1}
\end{equation}
As a map $x_{n+1}=F(x_{n})$ in Eq.~(\ref{eq:map0}) we take the chaotic driving
by a tent map 
\begin{equation}
x_{n+1}=\begin{cases}
2x_{n}\,, & 0\leq x_{n}\leq\frac{1}{2}\\
2-2x_{n}\,, & \frac{1}{2}\leq x_{n}\leq1.
\end{cases}\label{eq:ex-map-x}
\end{equation}
The variable $z_{n}$ with given average $\langle z\rangle$ and variance
$\langle(z-\langle z\rangle)^{2}\rangle$ can be obtained from $x_{n}$ using the
equation 
\begin{equation}
z_{n}=\sqrt{\frac{\langle(z-\langle z\rangle)^{2}\rangle}{\langle(x-\langle x\rangle)^{2}\rangle}}
(x_{n}-\langle x\rangle)+\langle z\rangle\,.
\label{eq:ex-map-z0}
\end{equation}
For the tent map (\ref{eq:ex-map-x}) the average and the variance are $\langle x\rangle=0.5$ 
and $\langle(x-\langle x\rangle)^{2}\rangle=1/12$, respectively. 

Another example is when the variable $z_{n}$ acquires only two values
$\pm\zeta$, with the probabilities $p_{+}$ and $p_{-}$, $p_{+}+p_{-}=1$. In
this case the average and the variance of $z_{n}$ are given by the
equations$\langle z\rangle=\zeta(p_{+}-p_{-})$ and $\langle(z-\langle
z\rangle)^{2}\rangle=4\zeta^{2}p_{+}p_{-}$. Expressing the probabilities we get 
\begin{equation}
p_{\pm}=\frac{1}{2}\pm\frac{\langle z\rangle}{2\sqrt{\langle(z-\langle z\rangle)^{2}\rangle
+\langle z\rangle^{2}}}
\label{eq:ppm}
\end{equation}
and
\begin{equation}
\zeta=\sqrt{\langle(z-\langle z\rangle)^{2}\rangle+\langle z\rangle^{2}}\,.
\label{eq:dzeta}
\end{equation}
In particular, if $\langle(z-\langle z\rangle)^{2}\rangle=1$, $\langle
z\rangle=5\times10^{-5}$ then $p_{+}\approx0.500025$, $p_{-}\approx0.499975$,
$\zeta\approx1.00000000125$.  Such two-valued variable $z_{n}$ can be
implemented by the following map:
\begin{equation}
y_{n+1}=\begin{cases}
y_{n}-y_{n}^{2}\zeta+0.5y_{n}^{3}\,, & 0\leqslant x_{n}\leqslant p_{-},\\
0.5-|y_{n}+y_{n}^{2}\zeta+0.5y_{n}^{3}-0.5|\,, & p_{-}<x_{n}\leqslant1. 
\end{cases}
\label{eq:ex-map-2}
\end{equation}
Note, that also in the map (\ref{eq:ex-map-2}) we use a reflection
at $y=0.5$ as a mechanism of reinjection. 

\begin{figure*}
\includegraphics[width=0.33\textwidth]{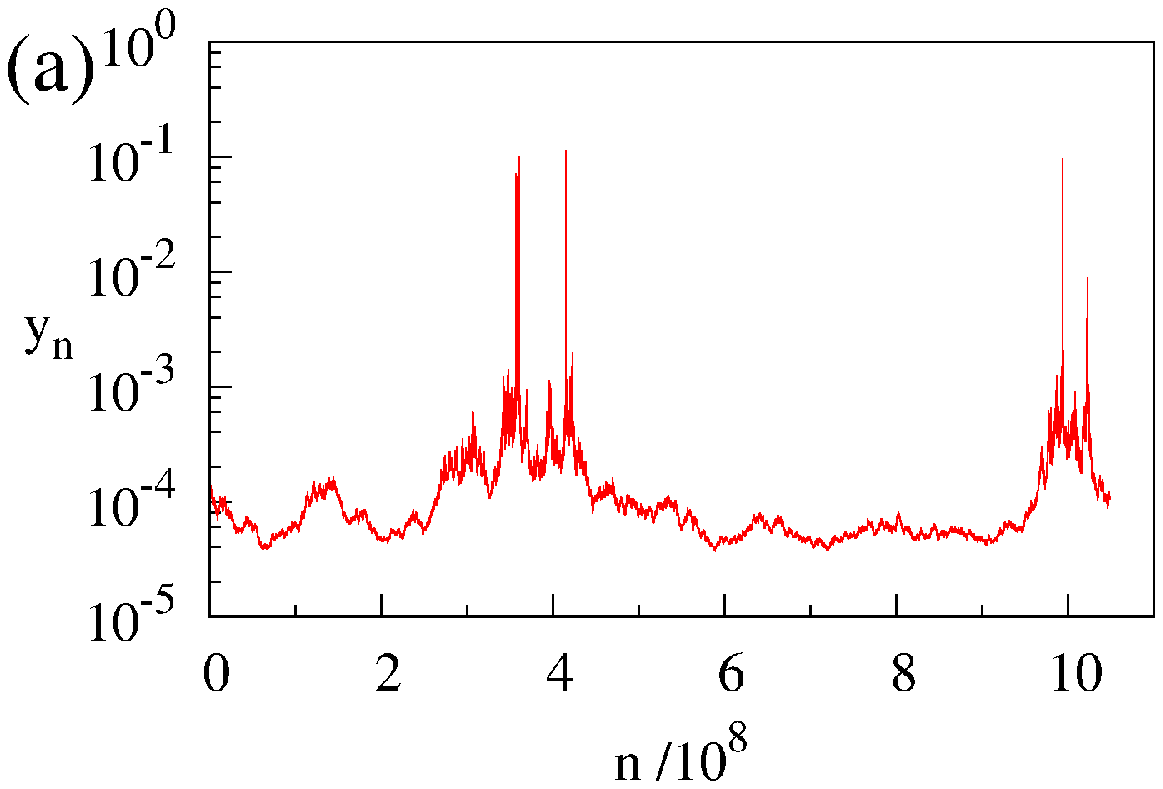}\includegraphics[width=0.33\textwidth]{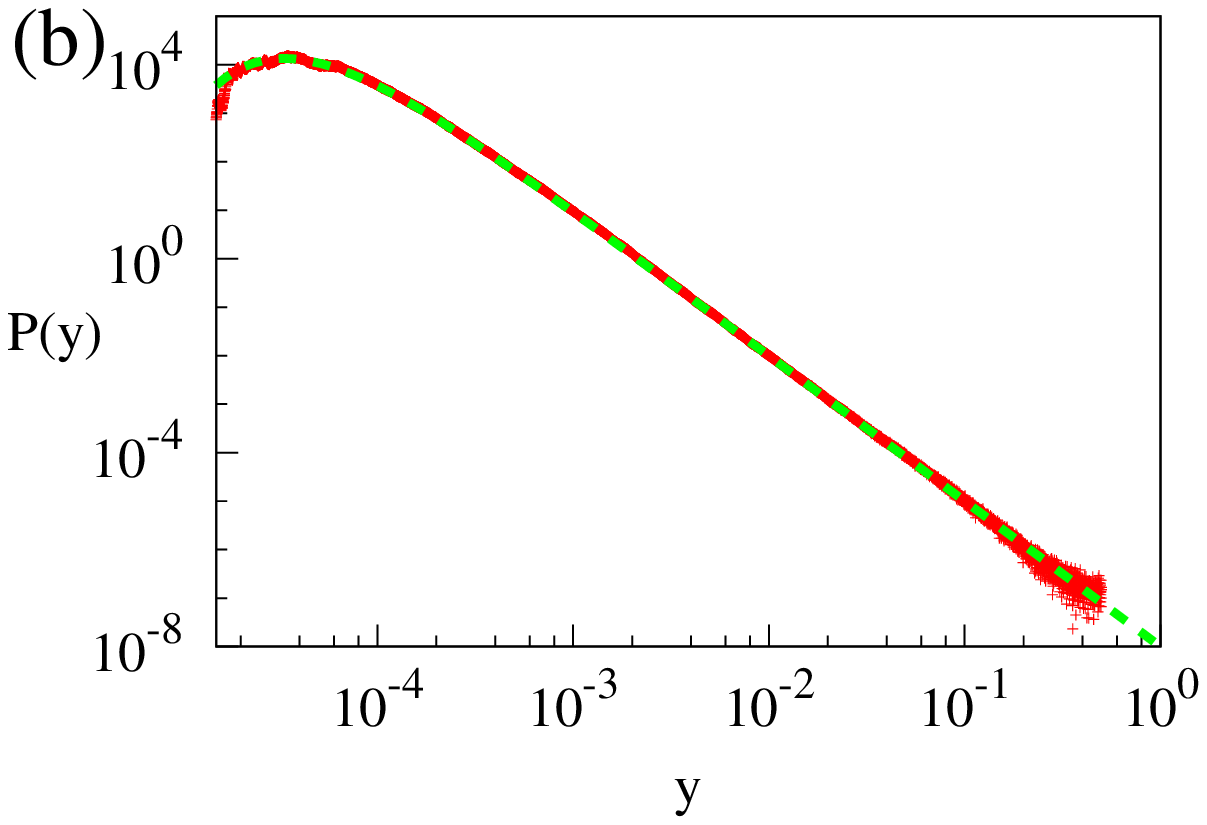}\includegraphics[width=0.33\textwidth]{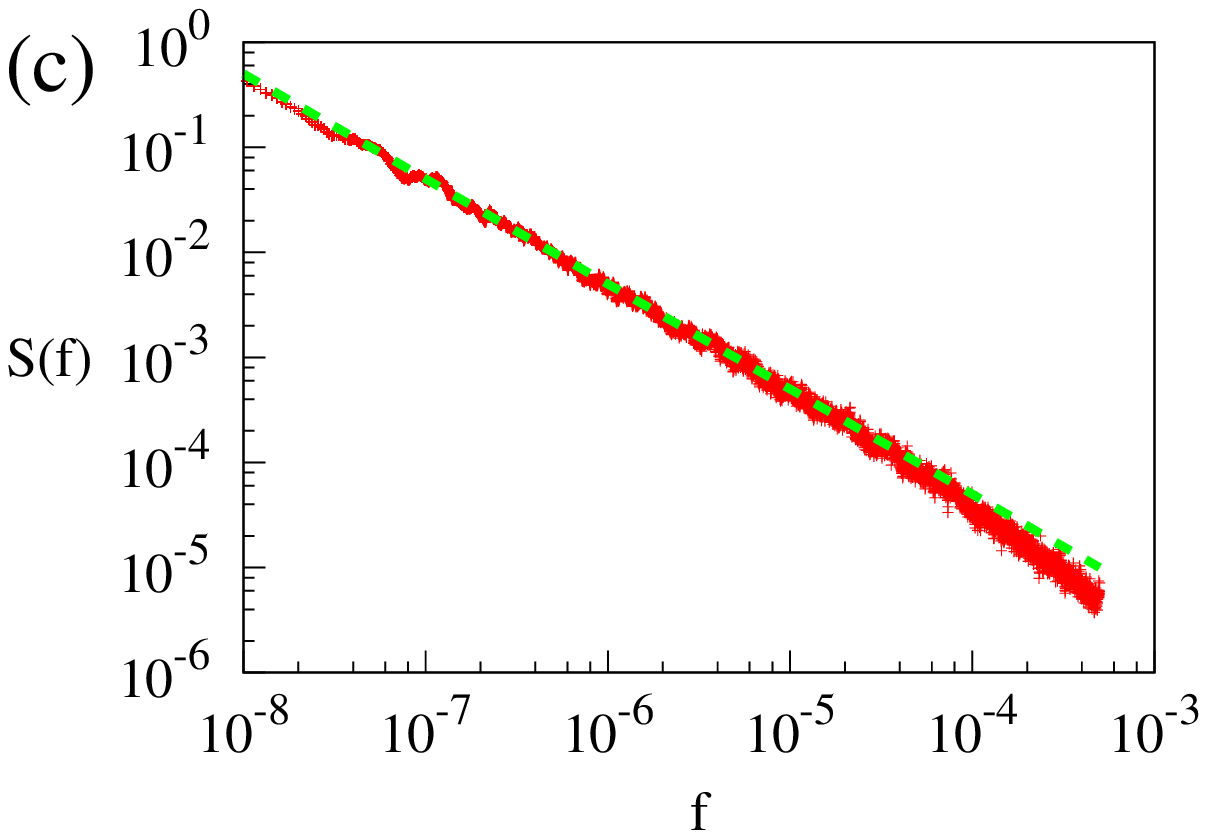}
\caption{(Color online) (a) Time series of map described by
  Eqs.~(\ref{eq:ex-map-1}), (\ref{eq:ex-map-x}), (\ref{eq:ex-map-z0}). (b) PDF
  of the variable $y$. The dashed (green) line is the analytical expression
  (\ref{eq:pdf-steady}).  (c) Power spectral density $S(f)$ of time series. The
  dashed (green) line shows the slope $1/f$. Parameters used are $\langle
z\rangle=5\times10^{-5}$ , $\langle(z-\langle z\rangle)^{2}\rangle=1$.}
\label{fig:map1}
\end{figure*}

\begin{figure*}
\includegraphics[width=0.33\textwidth]{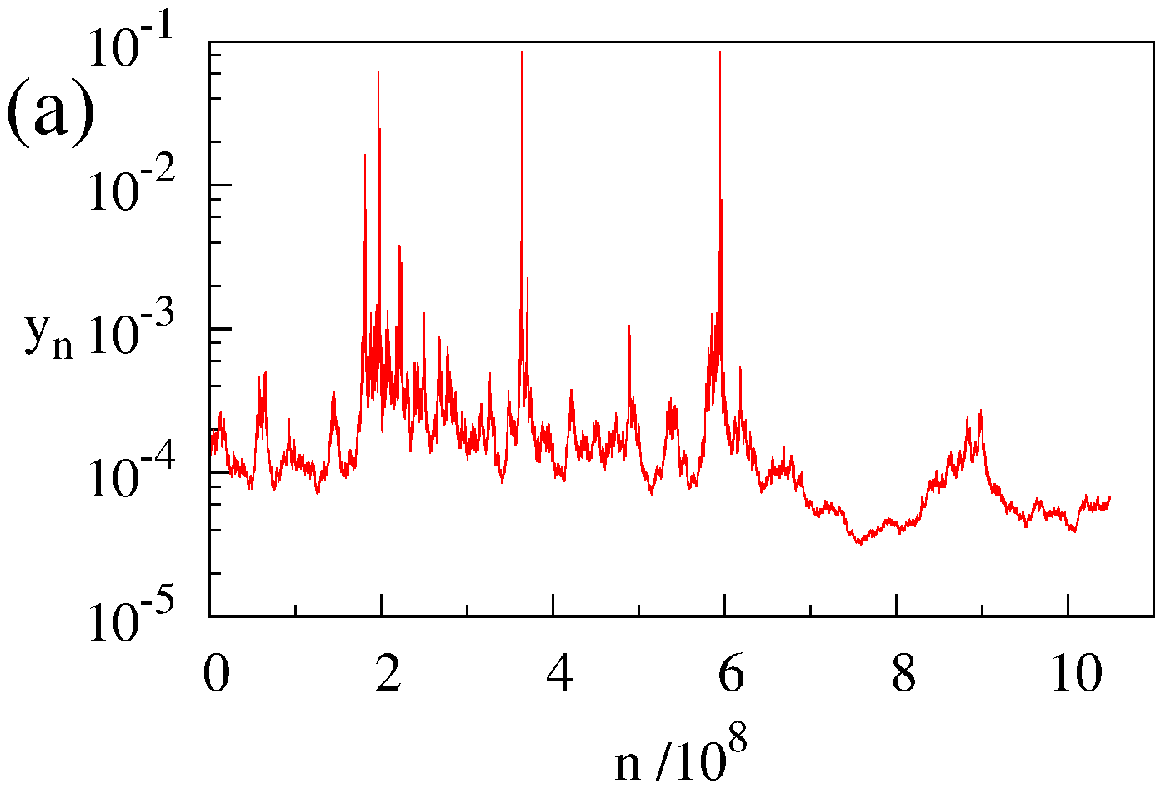}\includegraphics[width=0.33\textwidth]{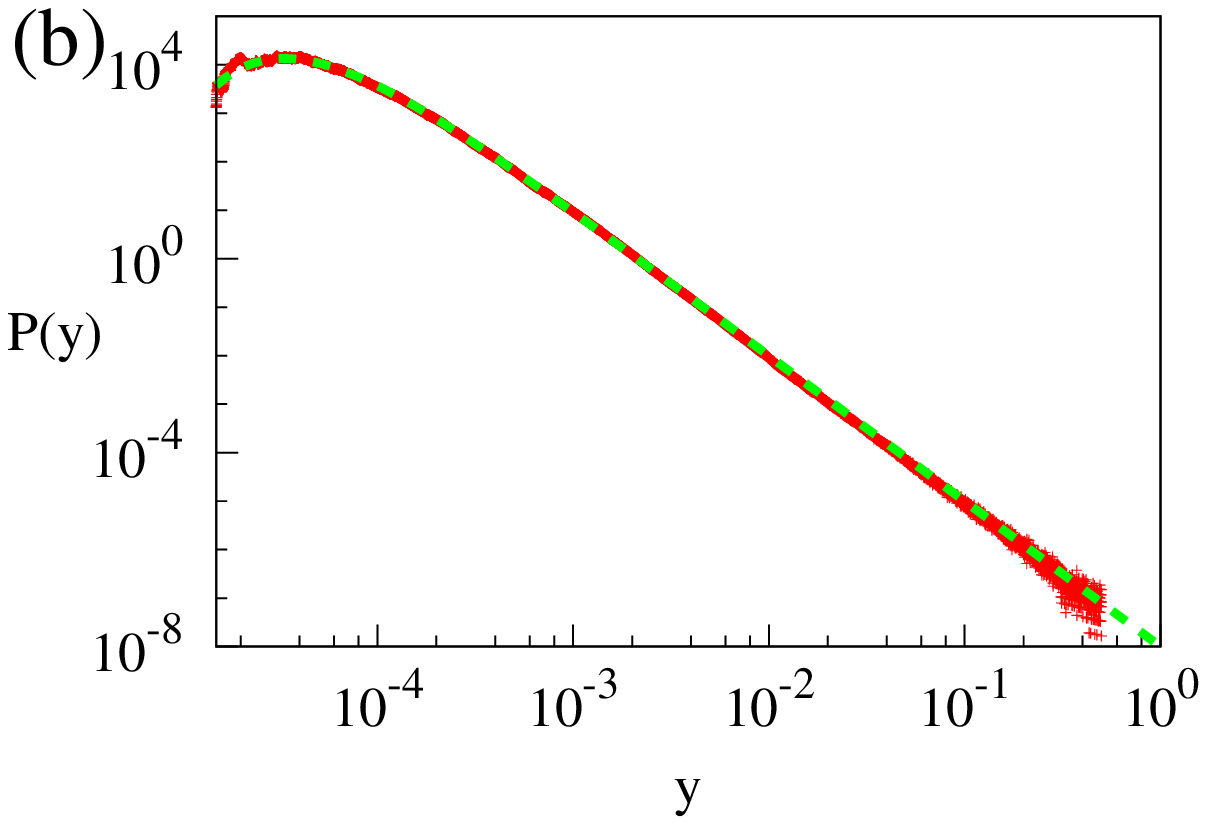}\includegraphics[width=0.33\textwidth]{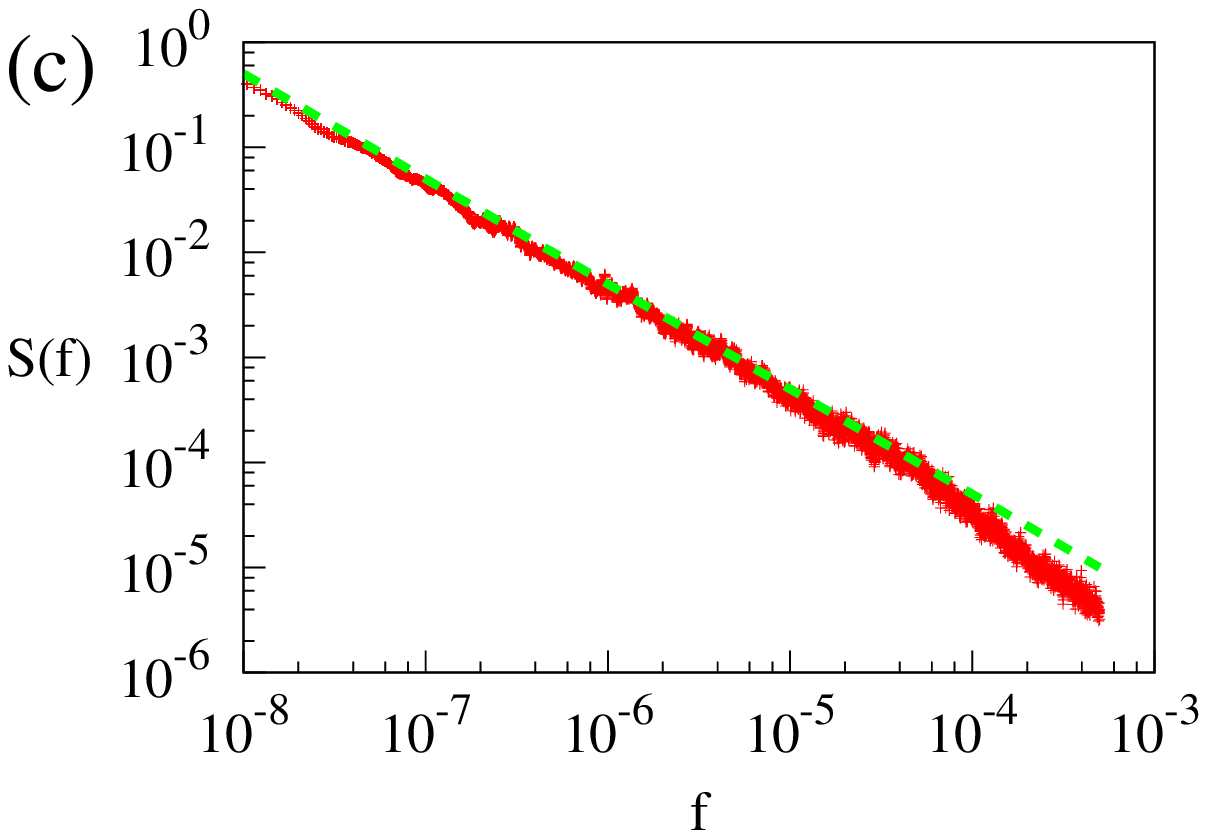}
\caption{(Color online) (a) Time series of map described by
  Eqs.~(\ref{eq:ex-map-x}), (\ref{eq:ppm})--(\ref{eq:ex-map-2}). (b) PDF of the
  variable $y$.  The dashed (green) line is the analytical expression
  (\ref{eq:pdf-steady}).  (c) Power spectral density $S(f)$ of time series. The
  dashed (green) line shows the slope $1/f$. Parameters used are the same as in
Fig.~\ref{fig:map1}.}
\label{fig:map2}
\end{figure*}

The numerical results for maps described by Eqs.~(\ref{eq:ex-map-1}),
(\ref{eq:ex-map-x}), (\ref{eq:ex-map-z0}) and by Eqs.~(\ref{eq:ex-map-x}),
(\ref{eq:ppm})--(\ref{eq:ex-map-2}) are shown in Fig.~\ref{fig:map1}
and Fig.~\ref{fig:map2}, respectively.
We calculate the power spectral density directly, according to the definition, as 
the normalized squared modulus of the Fourier transform of the signal, 
\begin{equation}
S(f)=\left\langle \frac{2}{N}\left|\sum_{n=1}^{N}y_{n}e^{-i2\pi fn}\right|^{2}\right\rangle, 
\end{equation}
where the angle brackets $\langle\cdot\rangle$ denote averaging over
realizations.  We used the time series $\{y_{n}\}$ of the length $N=10^9$ and
averaged over $100$ realizations with randomly chosen initial value $y_0$. 

From Fig.~\ref{fig:map1}a and Fig.~\ref{fig:map2}a we can see that these maps indeed lead
to intermittent behavior, where the laminar phases are changed by bursts
of activity corresponding to the large deviations of the variable
$y$ from the average value. The laminar phases of the first map appear
smoother than laminar phases of the second. The PDF of the variable
$y$, shown in Fig.~\ref{fig:map1}b and Fig.~\ref{fig:map2}b, has
in both cases a power-law form with the exponent $-3$ for larger values
of $y$, whereas for small values of $y$ the PDF decreases exponentially.
The PSD of the time series $\{y_{n}\}$, shown in Fig.~\ref{fig:map1}c
and Fig.~\ref{fig:map2}c, has $1/f$ behavior for a wide range of
frequencies. The $1/f$ interval in the PSD in Figs.~\ref{fig:map1}c,
\ref{fig:map2}c is $10^{-8}\lesssim f\lesssim10^{-4}$.

For the map (\ref{eq:map-q-exp}) we consider the case with $\eta=3$.
To avoid reaching of the limiting value $s=1/(\eta-1)=0.5$ we modify
the map (\ref{eq:map-on-off-log}) by introducing the reflection from
the boundary $s_{n}=0.5$: 
\begin{equation}
s_{n+1}=0.5-|s_{n}+z_{n}-0.5|\,.
\end{equation}
Then the map (\ref{eq:map-q-exp}) for the transformed variable
$y_{n}=\exp_{\eta}(s_{n})$ takes the form
\begin{equation}
y_{n+1}=\frac{1}{\sqrt{\left|\frac{1}{y_{n}^{2}}-2z_{n}\right|}}\,.
\label{eq:ex-map-q-exp}
\end{equation}
For the variable $x_{n}$ we again use the tent map (\ref{eq:ex-map-x}) and
calculate $z_{n}$ according to Eq.~(\ref{eq:ex-map-z0}) with the the average
$\langle z\rangle=9\times10^{-4}$ and the variance$\langle(z-\langle
z\rangle)^{2}\rangle=1$.

\begin{figure*}
\includegraphics[width=0.33\textwidth]{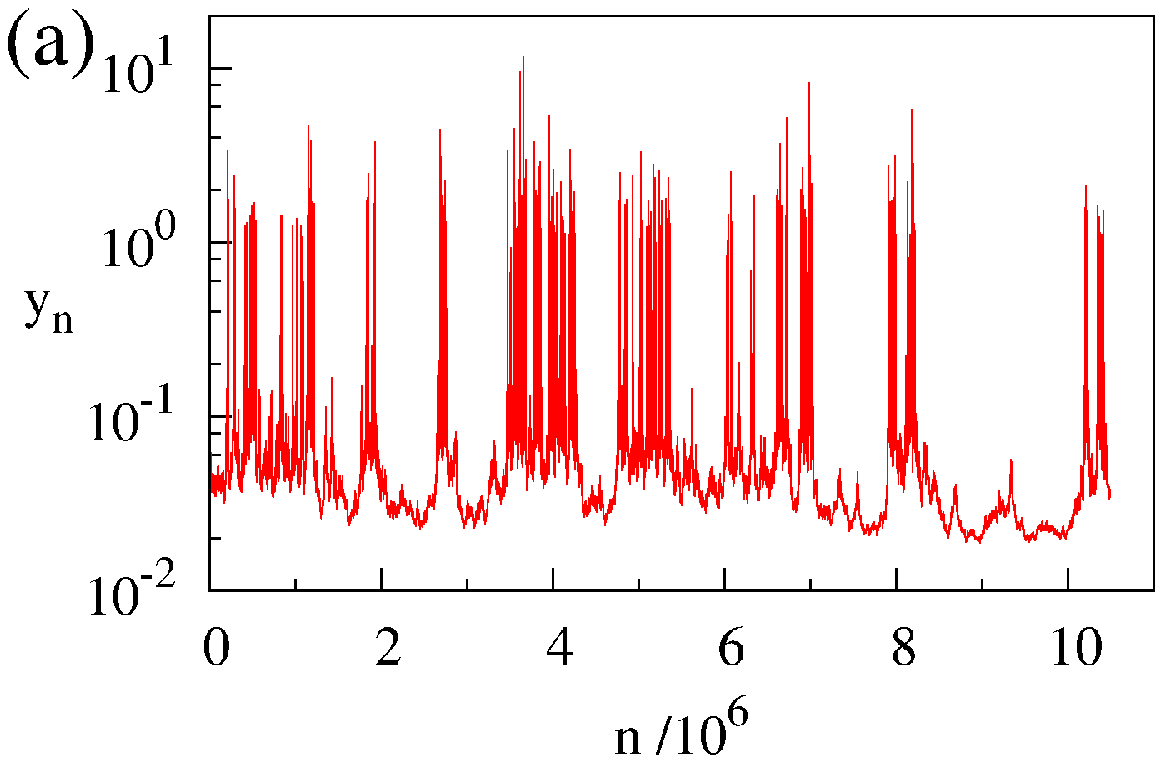}\includegraphics[width=0.33\textwidth]{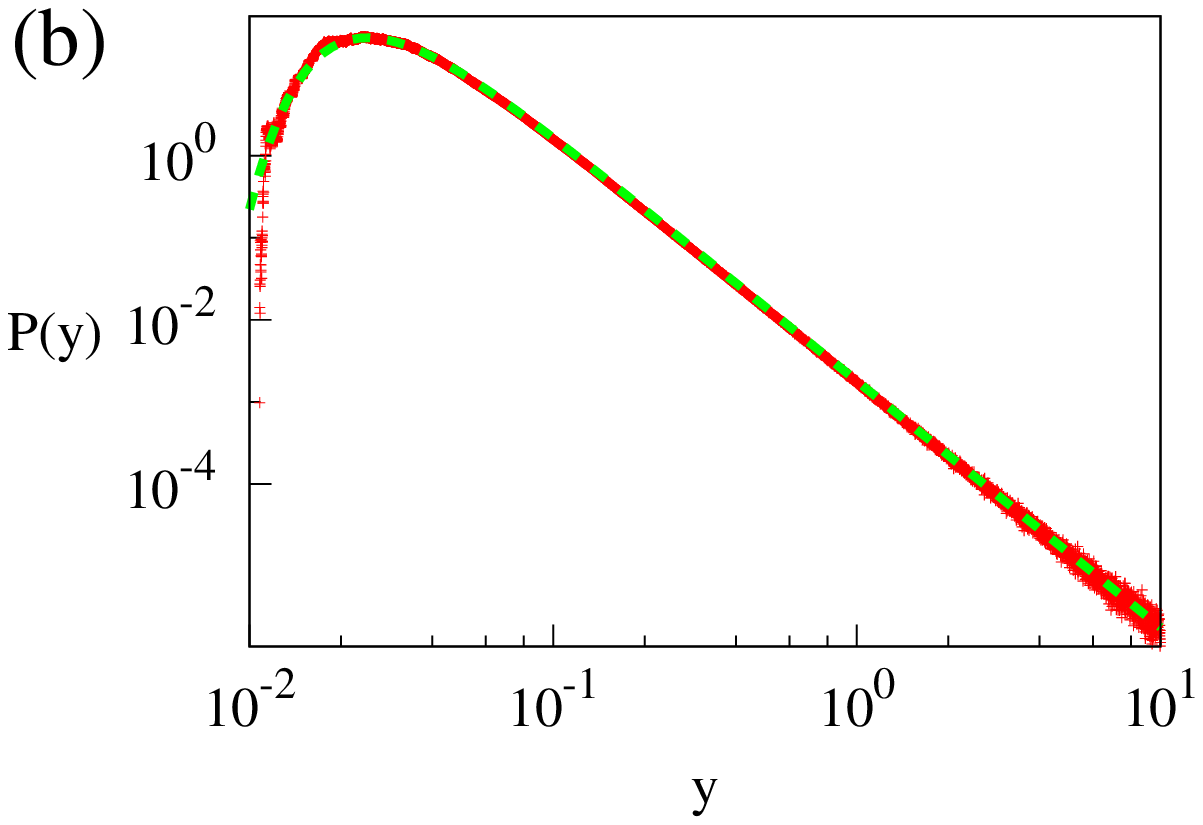}\includegraphics[width=0.33\textwidth]{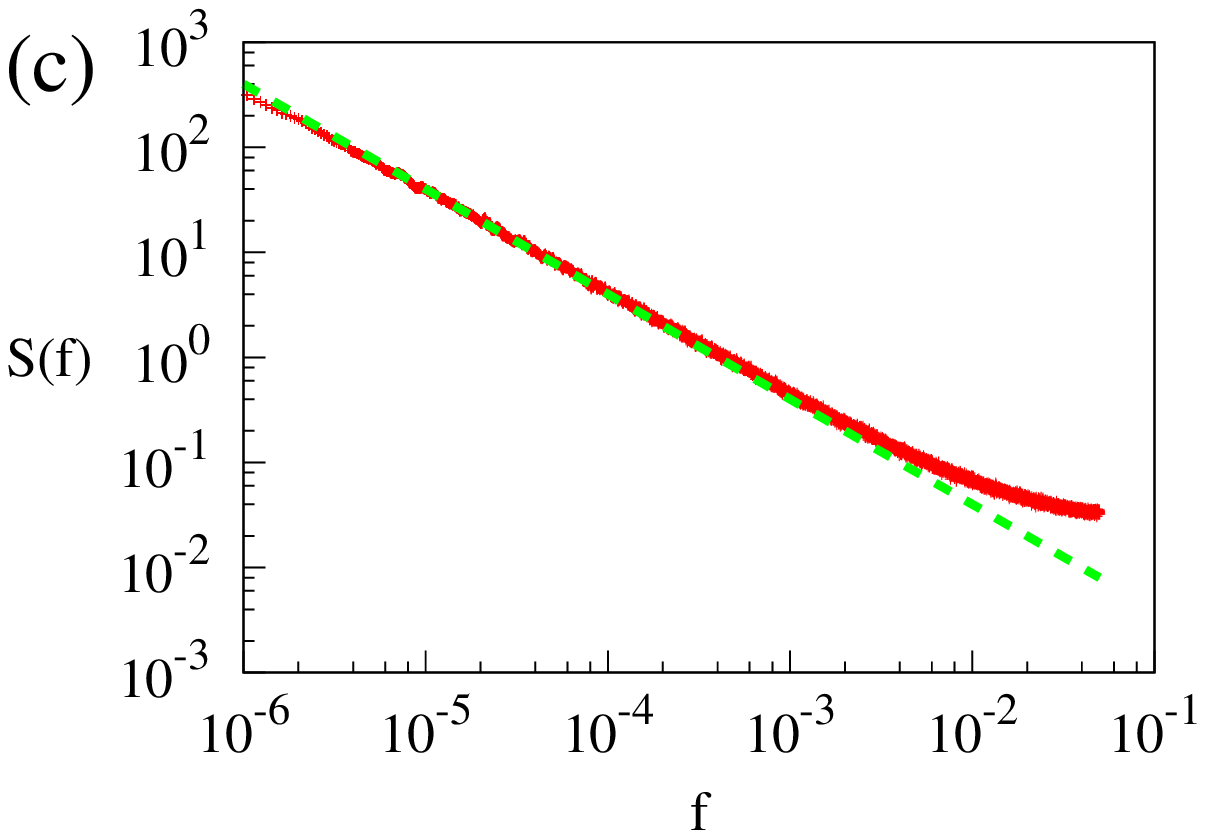}
\caption{(Color online) (a) Time series of map described by
  Eqs.~(\ref{eq:ex-map-x}), (\ref{eq:ex-map-z0}), (\ref{eq:ex-map-q-exp}). (b)
  PDF of the variable $y$. The dashed (green) line is the analytical expression
  (\ref{eq:pdf-steady}).  (c) Power spectral density $S(f)$ of time series. The
  dashed (green) line shows the slope $1/f$. Parameters used are $\langle
z\rangle=9\times10^{-4}$, $\langle(z-\langle z\rangle)^{2}\rangle=1$.}
\label{fig:map3}
\end{figure*}

The numerical results for the map described by Eqs.~(\ref{eq:ex-map-x}),
(\ref{eq:ex-map-z0}), (\ref{eq:ex-map-q-exp}) are shown in Fig.~\ref{fig:map3}.
From Fig.~\ref{fig:map3}a we can see that this map leads to the intermittent
behavior. Due to larger average $\langle z\rangle$ and larger exponent
$\eta$ the durations of laminar phases are shorter than in Figs.~\ref{fig:map1}a,
\ref{fig:map2}a. The PDF of the variable $y$, shown in Fig.~\ref{fig:map3}b,
has a power-law form with the exponent $-3$ for larger values of $y$,
whereas for small values of $y$ the PDF decreases exponentially.
The PSD of the time series $\{y_{n}\}$, shown in Fig.~\ref{fig:map3}c,
has $1/f$ behavior for a wide range of frequencies. The $1/f$ interval
in the PSD is $10^{-6}\lesssim f\lesssim10^{-3}$.

As the numerical examples show, both maps (\ref{eq:map2}) and
(\ref{eq:map-q-exp}) for some values of the parameters can yield time series
with $1/f$ PSD in a wide range of frequencies. In addition, the PDF of the
deviation from the invariant subspace $y$ for these values of parameters has a
power-law part with the exponent $-3$, in contrast to on-off intermittency
where the exponent in the PDF is close to $-1$ and $1/\sqrt{f}$ PSD.  The
explanation of the observed behavior of PDF and PSD will be provided in the next
Section. 

\section{\label{sec:map-SDE}Approximation of discrete maps by stochastic differential equations}

To obtain analytical expressions for the PDF and PSD of the
deviation $y$, we approximate the maps (\ref{eq:map2}) and (\ref{eq:map-q-exp})
by a SDE. To obtain the SDE corresponding to the map (\ref{eq:map2}) we proceed
as follows: we replace the variable $z_{n}$ by a random Gaussian variable
having the same average and variance as $z_{n}$ and interpret
Eq.~(\ref{eq:map2}) as Euler-Marujama approximation of a SDE. In this way we
get the following SDE:
\begin{equation}
dy=\sigma^{2}\left(\eta-\frac{\nu}{2}+
\frac{\eta-1}{2}\left(\frac{y_{\mathrm{min}}}{y}\right)^{\eta-1}\right)y^{2\eta-1}dt
+\sigma y^{\eta}dW\,.
\label{eq:sde-map}
\end{equation}
Here $W$ is a standard Wiener process (the Brownian motion) and the parameters
$\sigma$, $y_{\mathrm{min}}$, and $\nu$ are given by the equations 
\begin{eqnarray}
\sigma & = & \sqrt{\langle(z-\langle z\rangle)^{2}\rangle}\,,\label{eq:sigma}\\
y_{\mathrm{min}} & = & \left[\frac{2\langle z\rangle}{(\eta-1)\langle(z-\langle z\rangle)^{2}\rangle}
\right]^{\frac{1}{\eta-1}}\,,\label{eq:ymin}\\
\nu & = & 2\eta-\frac{2\gamma}{\langle(z-\langle z\rangle)^{2}\rangle}\,.\label{eq:nu-z}
\end{eqnarray}
SDE approximating the map (\ref{eq:map-q-exp}) can be obtained in the following
way: we approximate a random walk described by Eq.~(\ref{eq:map-on-off-log}) by
a Brownian motion with constant drift $ds=adt+\sigma dW$, where $\sigma$ is
given by Eq.~(\ref{eq:sigma}) and $a=\langle z\rangle$.  After transformation
of the variable $s$ to the variable $y=\exp_{\eta}(s)$ we get a particular case
of Eq.~(\ref{eq:sde-map}) with $\nu=\eta$, the other parameters $\sigma$ and
$y_{\mathrm{min}}$ are given by Eqs.~(\ref{eq:sigma}), (\ref{eq:ymin}). Thus,
both maps (\ref{eq:map2}) and (\ref{eq:map-q-exp}) correspond to the same SDE
(\ref{eq:sde-map}). The SDE (\ref{eq:sde-map}) has the same form as that
considered in Refs.~\onlinecite{Kaulakys2004,Kaulakys2006}.
It is possible to obtain the non-linear SDE of the form (\ref{eq:sde-map})
starting from the agent-based herding model. \cite{ruseckas-2011epl}
In Ref.~\onlinecite{ruseckas-2011} modifications of these
equations by introducing additional parameters are presented. These equations
may generate signals with q-exponential and q-Gaussian distribution of the
nonextensive statistical mechanics.

The approximation of the map (\ref{eq:map2}) by the SDE (\ref{eq:sde-map}) is
valid when the value of $y$ is sufficiently small. The maximum value of $y$ can
be determined from the condition that the second term in Eq.~(\ref{eq:map2})
should be much smaller than the first.  We can estimate this condition as 
\begin{equation}
y_{\mathrm{max}}^{\eta}\sqrt{\langle(z-\langle z\rangle)^{2}\rangle}
\ll y_{\mathrm{max}}
\end{equation}
giving
\begin{equation}
y_{\mathrm{max}}\lesssim\langle(z-\langle z\rangle)^{2}\rangle^{-\frac{1}{2(\eta-1)}}\,.
\label{eq:ymax}
\end{equation}
For the map (\ref{eq:map-q-exp}) the approximation by the SDE
(\ref{eq:sde-map}) is valid as long as the variable $s_{n}$ is far from
$1/(\eta-1)$ where the $q$-exponential function $\exp_{\eta}(s_{n})$ becomes
infinite. Assuming that the presence of the limiting value $1/(\eta-1)$ does
not influence the random walk (\ref{eq:map-on-off-log}) when the distance to
this limiting value is larger than the standard deviation of $z_{n}$ (that is,
$s_{n}<1/(\eta-1)-\sigma$) we can estimate the maximum value of $y$ as
$y_{\mathrm{max}}\lesssim\exp_{\eta}(1/(\eta-1)-\sigma)$.  This estimation
coincides with Eq.~(\ref{eq:ymax}).

Using Eqs.~(\ref{eq:ymin}) and (\ref{eq:ymax}) we get the expression for the
ratio $y_{\mathrm{max}}/y_{\mathrm{min}}$ : 
\begin{equation}
\frac{y_{\mathrm{max}}}{y_{\mathrm{min}}}\lesssim
\left[\frac{(\eta-1)\sqrt{\langle(z-\langle z\rangle)^{2}\rangle}}{2\langle z\rangle}\right]^{\frac{1}{\eta-1}}\,.
\label{eq:ratio-min-max}
\end{equation}
As it was shown in Refs.~\onlinecite{Kaulakys2004,Kaulakys2006}, the SDE
(\ref{eq:sde-map}) generates signals with power-law PSD in a wide range of
frequencies when the variable $y$ can vary in a wide region,
$y_{\mathrm{max}}\gg y_{\mathrm{min}}$. The condition
$y_{\mathrm{max}}/y_{\mathrm{min}}\gg1$ is obeyed when
\begin{equation}
\langle(z-\langle z\rangle)^{2}\rangle\gg\langle z\rangle^{2}\,,
\end{equation}
that is, the standard deviation of the variable $z_{n}$ should be much larger
than the average.

The SDE (\ref{eq:sde-map}) leads to the steady state PDF 
\begin{equation}
P_{0}(y)=\frac{(\eta-1)y_{\mathrm{min}}^{\nu-1}}{\Gamma\left(\frac{\nu-1}{\eta-1}\right)y^{\nu}}
\exp\left[-\left(\frac{y_{\mathrm{min}}}{y}\right)^{\eta-1}\right]\,.
\label{eq:pdf-steady}
\end{equation}
Thus, the parameter $\nu$ gives the exponent of the power-law part of the PDF
and the parameter $y_{\mathrm{min}}$ gives the position of the exponential
cut-off at small values of $y$. From Eq.~(\ref{eq:ymin}) it follows that
$y_{\mathrm{min}}$ grows with the growing average $\langle z\rangle$.  As can
be seen in Figs.~\ref{fig:map1}b, \ref{fig:map2}b, \ref{fig:map3}b, there is a
good agreement of the numerically obtained PDF with the analytical expression
(\ref{eq:pdf-steady}). Similarly as in the case of on-off intermittency we
obtain PDF of the deviation from the invariant subspace having power-law form,
however, the power exponent $\nu$ can assume values significantly different
from $1$. 

Numerical analysis, performed in Ref.~\onlinecite{Kaulakys2009} indicates that the
stochastic variable $y$, described by a SDE similar to (\ref{eq:sde-map})
exhibits intermittent behavior: there are peaks, bursts or extreme events,
corresponding to the large deviations of the variable from the appropriate
average value, separated by laminar phases with a wide range distribution of
the laminar durations. The exponent $-3/2$ in the PDF of the interburst
durations has been numerically obtained.

In Refs.~\onlinecite{Kaulakys2004,Kaulakys2006} it was shown that SDE
(\ref{eq:sde-map}) generates signals with PSD having the form $S(f)\sim
f^{-\beta}$ in a wide range of frequencies with the exponent
\begin{equation}
\beta=1+\frac{\nu-3}{2(\eta-1)}\,.
\label{eq:beta}
\end{equation}
The connection of the PSD of the signal generated by SDE (\ref{eq:sde-map})
with the behavior of the eigenvalues of the corresponding Fokker-Planck
equation was analyzed in Ref.~\onlinecite{Ruseckas10}. An additional argument based
on scaling properties showing that PSD of the signal generated by
SDE~(\ref{eq:sde-map}) has the power-law behavior in some range of frequencies
we present in Appendix~\ref{sec:SDE}. For the parameters used in
Figs.~\ref{fig:map1}, \ref{fig:map2}, \ref{fig:map3}, Eq.~(\ref{eq:beta}) gives
$\beta=1$. Numerically obtained PSD shown in Figs.~\ref{fig:map1}c,
\ref{fig:map2}c, \ref{fig:map3}c confirms this prediction. Thus, as long as the
approximation of the maps (\ref{eq:map2}) or (\ref{eq:map-q-exp}) by the
SDE~(\ref{eq:sde-map}) is valid, the PSD of the time series $\{y_{n}\}$
exhibits a power-law behavior, including $1/f$ noise.

The range of frequencies where PSD has power-law behavior is limited by the
minimum and maximum values $y_{\mathrm{min}}$ and $y_{\mathrm{max}}$.  The
limiting frequencies are estimated in Ref.~\onlinecite{Ruseckas10} and also in
Appendix~\ref{sec:append-B}. Using Eqs.~(\ref{eq:sigma}), (\ref{eq:ymin}) and
(\ref{eq:ymax}), we can write the range of frequencies (\ref{eq:freq-range})
where the PSD has the power-law form as
\begin{equation}
\left(\frac{y_{\mathrm{min}}}{y_{\mathrm{max}}}\right)^{2(\eta-1)}\ll2\pi f\ll1\,.
\label{eq:freq-range-2}
\end{equation}
If $y_{\mathrm{max}}/y_{\mathrm{min}}\gg1$, this frequency range can span many
orders of magnitude. However, this estimation of the frequency range is too
broad and the numerical solution of Eq.~(\ref{eq:sde-map}) gives much narrower
range. Nevertheless, Eq.~(\ref{eq:freq-range-2}) correctly reflects the
following properties of the frequency region where PSD has $1/f^{\beta}$
dependence: the width of this frequency region increases with increase of the
ratio between minimum and maximum values, $y_{\mathrm{min}}$ and
$y_{\mathrm{max}}$, and with increase of the difference $\eta-1$.
\cite{Ruseckas10}

\section{\label{sec:concl}Conclusions}

We demonstrate that the nonlinear maps having invariant
subspace and the expansion in the powers of the deviation from the invariant
subspace having the form of Eq.~(\ref{eq:map1}) can generate signals with $1/f$
noise. In contrast to known mechanism of $1/f$ noise involving
Pomeau-Manneville type maps, the parameter $z_{n}$ in the map
Eq.~(\ref{eq:map1}) is not static. Another difference is that the exponent
$\beta$ in the PSD, as Eq.~(\ref{eq:beta}) shows, depends on two parameters
$\eta$ and $\nu$, thus $1/f^\beta$ noise can be obtained for various values of
the exponent $\beta$. 

The width of the frequency region where the PSD has $f^{-\beta}$ behavior is
limited by the average value of the variable $z_{n}$: this width increases as
$\langle z\rangle$ approaches the threshold value $\langle z\rangle=0$. In
addition, the width of the power-law region in the PSD increases with
increasing the difference $\eta-1$. 

\appendix

\section{\label{sec:SDE}Nonlinear stochastic differential equation generating signals with
$1/f^{\beta}$ noise}

Pure $1/f^{\beta}$ PSD is physically impossible because the total power would
be infinity. Therefore we will consider signals with PSD having $1/f^{\beta}$
behavior only in some wide intermediate region of frequencies,
$f_{\mathrm{min}}\ll f\ll f_{\mathrm{max}}$, whereas for small frequencies
$f\ll f_{\mathrm{min}}$ PSD is bounded. We can obtain nonlinear SDE generating
signals exhibiting $1/f$ noise using the following considerations.
Wiener-Khintchine theorem relates PSD $S(f)$ to the autocorrelation function
$C(t)$:
\begin{equation}
C(t)=\int_{0}^{+\infty}S(f)\cos(2\pi ft)dt\,.\label{eq:W-K}
\end{equation}
If $S(f)\sim f^{-\beta}$ in a wide region of frequencies, then for the
frequencies in this region the PSD has a scaling property
\begin{equation}
S(af)\sim a^{-\beta}S(f)
\end{equation}
when the influence of the limiting frequencies $f_{\mathrm{min}}$ an
$f_{\mathrm{max}}$ is neglected. From the Wiener-Khintchine theorem
(\ref{eq:W-K}) it follows that the autocorrelation function has the scaling
property
\begin{equation}
C(at)\sim a^{\beta-1}C(t)\label{eq:auto-scaling}
\end{equation}
in the time range $1/f_{\mathrm{max}}\ll t\ll1/f_{\mathrm{min}}$.  The
autocorrelation function can be written as
\cite{Ruseckas10,Risken1996,Gardiner04}
\begin{equation}
C(t)=\int dy\int dy'\, yy'P_{0}(y)P_{y}(y',t|y,0)\,,\label{eq:auto}
\end{equation}
where $P_{0}(y)$ is the steady-state PDF and $P_{y}(y',t|y,0)$ is the
transition probability (the conditional probability that at time $t$ the signal
has value $y'$ with the condition that at time $t=0$ the signal had the value
$y$). The transition probability can be obtained from the solution of the
Fokker-Planck equation with the initial condition
$P_{y}(y',t|y,0)=\delta(y'-y)$. The required property (\ref{eq:auto-scaling})
can be obtained when the steady-state PDF has the power-law form
\begin{equation}
P_{0}(y)\sim y^{-\nu}\label{eq:prob-steady}
\end{equation}
and the transition probability has the scaling property
\begin{equation}
P_{y}(ay',t|ay,0)=a^{-1}P_{y}(y',a^{2(\eta-1)}t|y,0)\,,\label{eq:prob-scaling}
\end{equation}
that is, change of the magnitude of the stochastic variable $y$ is equivalent
to the change of time scale. In this case from Eq.~(\ref{eq:auto}) it follows
that the autocorrelation function has the required property
(\ref{eq:auto-scaling}) with $\beta$ given by Eq.~(\ref{eq:beta}).  In order to
avoid the divergence of steady state PDF (\ref{eq:prob-steady}) the diffusion
of stochastic variable $y$ should be restricted at least from the side of small
values and, therefore, Eq.~(\ref{eq:prob-steady}) holds only in some region of
the variable $y$, $y_{\mathrm{min}}\ll y\ll y_{\mathrm{max}}$.  When the
diffusion of stochastic variable $y$ is restricted, Eq.~(\ref{eq:prob-scaling})
also cannot be exact. However, if the influence of the limiting values
$y_{\mathrm{min}}$ and $y_{\mathrm{max}}$ can be neglected for time $t$ in some
region $t_{\mathrm{min}}\ll t\ll t_{\mathrm{max}}$, we can expect that
Eq.~(\ref{eq:auto-scaling}) approximately holds for this time region.

To get the required scaling (\ref{eq:prob-scaling}) of the transition
probability, the SDE should contain only powers of the stochastic variable $y$
and the coefficient in the noise term should be proportional to $y^{\eta}$. The
drift term then is fixed by the requirement (\ref{eq:prob-steady}) for the
steady-state PDF. Thus we consider SDE 
\begin{equation}
dy=\sigma^{2}\left(\eta-\frac{1}{2}\nu\right)y^{2\eta-1}dt+\sigma y^{\eta}dW\,.\label{eq:sde}
\end{equation}
In order to obtain a stationary process and avoid the divergence of steady
state PDF the diffusion of stochastic variable $y$ should be restricted or
equation (\ref{eq:sde}) should be modified. The simplest choice of the
restriction is the reflective boundary conditions at $y=y_{\mathrm{min}}$ and
$y=y_{\mathrm{max}}$. Exponentially restricted diffusion with the steady state
PDF 
\begin{equation}
P_{0}(y)\sim\frac{1}{y^{\nu}}\exp\left\{ -\left(\frac{y_{\mathrm{min}}}{y}\right)^{m}
-\left(\frac{y}{y_{\mathrm{max}}}\right)^{m}\right\} 
\end{equation}
is generated by the SDE
\begin{equation}
dy=\sigma^{2}\left[\eta-\frac{1}{2}\nu+\frac{m}{2}\left(\frac{y_{\mathrm{min}}^{m}}{y^{m}}
-\frac{y^{m}}{y_{\mathrm{max}}^{m}}\right)\right]y^{2\eta-1}dt+\sigma y^{\eta}dW
\label{eq:sde-restricted}
\end{equation}
obtained from Eq.~(\ref{eq:sde}) by introducing the additional terms.

\section{\label{sec:append-B}Estimation of the range of the frequencies where
PSD has the power-law behavior}

The presence of the restrictions at $y=y_{\mathrm{min}}$
and $y=y_{\mathrm{max}}$ makes the scaling (\ref{eq:prob-scaling}) not exact
and this limits the power-law part of the PSD to a finite range of frequencies
$f_{\mathrm{min}}\ll f\ll f_{\mathrm{max}}$.  Let us estimate the limiting
frequencies. Taking into account the limiting values $y_{\mathrm{min}}$ and
$y_{\mathrm{max}}$, Eq.~(\ref{eq:prob-scaling}) for the transition probability
corresponding to SDE (\ref{eq:sde}) becomes
\begin{multline}
P_{y}(ay',t|ay,0;ay_{\mathrm{min}},ay_{\mathrm{max}})=\\
a^{-1} P_{y}(y',a^{2(\eta-1)}t|y,0;y_{\mathrm{min}},y_{\mathrm{max}})\,.
\label{eq:prob-scaling-bound}
\end{multline}
The steady-state distribution $P_{0}(y;y_{\mathrm{min}},y_{\mathrm{max}})$ has
the scaling property
\begin{equation}
P_{0}(ay;ay_{\mathrm{min}},ay_{\mathrm{max}})=
a^{-1}P_{0}(y;y_{\mathrm{min}},y_{\mathrm{max}})\,.
\label{eq:prob-steady-bound}
\end{equation}
Inserting Eqs.~(\ref{eq:prob-scaling-bound}) and (\ref{eq:prob-steady-bound})
into Eq.~(\ref{eq:auto}) we obtain
\begin{equation}
C(t;ay_{\mathrm{min}},ay_{\mathrm{max}})=a^{2}C(a^{2(\eta-1)}t,y_{\mathrm{min}},y_{\mathrm{max}})\,.
\end{equation}
This equation means that time $t$ in the autocorrelation function should enter
only in combinations with the limiting values,
$y_{\mathrm{min}}t^{\frac{1}{2(\eta-1)}}$ and
$y_{\mathrm{max}}t^{\frac{1}{2(\eta-1)}}$. We can expect that the influence of
the limiting values can be neglected and Eq.~(\ref{eq:prob-scaling}) holds when
the first combination is small and the second large, that is when time $t$ is
in the interval $\sigma^{-2}y_{\mathrm{max}}^{2(1-\eta)}\ll
t\ll\sigma^{-2}y_{\mathrm{min}}^{2(1-\eta)}$.  Then, using Eq.~(\ref{eq:W-K})
the frequency range where the PSD has $1/f^{\beta}$ behavior can be estimated
as
\begin{equation}
\sigma^{2}y_{\mathrm{min}}^{2(\eta-1)}\ll2\pi f\ll
\sigma^{2}y_{\mathrm{max}}^{2(\eta-1)}\,.
\label{eq:freq-range}
\end{equation}


\begin{thebibliography}{69}%
\makeatletter
\providecommand \@ifxundefined [1]{%
 \@ifx{#1\undefined}
}%
\providecommand \@ifnum [1]{%
 \ifnum #1\expandafter \@firstoftwo
 \else \expandafter \@secondoftwo
 \fi
}%
\providecommand \@ifx [1]{%
 \ifx #1\expandafter \@firstoftwo
 \else \expandafter \@secondoftwo
 \fi
}%
\providecommand \natexlab [1]{#1}%
\providecommand \enquote  [1]{``#1''}%
\providecommand \bibnamefont  [1]{#1}%
\providecommand \bibfnamefont [1]{#1}%
\providecommand \citenamefont [1]{#1}%
\providecommand \href@noop [0]{\@secondoftwo}%
\providecommand \href [0]{\begingroup \@sanitize@url \@href}%
\providecommand \@href[1]{\@@startlink{#1}\@@href}%
\providecommand \@@href[1]{\endgroup#1\@@endlink}%
\providecommand \@sanitize@url [0]{\catcode `\\12\catcode `\$12\catcode
  `\&12\catcode `\#12\catcode `\^12\catcode `\_12\catcode `\%12\relax}%
\providecommand \@@startlink[1]{}%
\providecommand \@@endlink[0]{}%
\providecommand \url  [0]{\begingroup\@sanitize@url \@url }%
\providecommand \@url [1]{\endgroup\@href {#1}{\urlprefix }}%
\providecommand \urlprefix  [0]{URL }%
\providecommand \Eprint [0]{\href }%
\providecommand \doibase [0]{http://dx.doi.org/}%
\providecommand \selectlanguage [0]{\@gobble}%
\providecommand \bibinfo  [0]{\@secondoftwo}%
\providecommand \bibfield  [0]{\@secondoftwo}%
\providecommand \translation [1]{[#1]}%
\providecommand \BibitemOpen [0]{}%
\providecommand \bibitemStop [0]{}%
\providecommand \bibitemNoStop [0]{.\EOS\space}%
\providecommand \EOS [0]{\spacefactor3000\relax}%
\providecommand \BibitemShut  [1]{\csname bibitem#1\endcsname}%
\let\auto@bib@innerbib\@empty
\bibitem [{\citenamefont {Batchelor}\ and\ \citenamefont
  {Townsend}(1949)}]{Batchelor1949}%
  \BibitemOpen
  \bibfield  {author} {\bibinfo {author} {\bibfnamefont {G.}~\bibnamefont
  {Batchelor}}\ and\ \bibinfo {author} {\bibfnamefont {A.}~\bibnamefont
  {Townsend}},\ }\href@noop {} {\bibfield  {journal} {\bibinfo  {journal}
  {Proc. R. Soc. London, Ser. A}\ }\textbf {\bibinfo {volume} {199}},\ \bibinfo
  {pages} {238} (\bibinfo {year} {1949})}\BibitemShut {NoStop}%
\bibitem [{\citenamefont {Pommeau}\ and\ \citenamefont
  {Manneville}(1980)}]{Pommeau1980}%
  \BibitemOpen
  \bibfield  {author} {\bibinfo {author} {\bibfnamefont {Y.}~\bibnamefont
  {Pommeau}}\ and\ \bibinfo {author} {\bibfnamefont {P.}~\bibnamefont
  {Manneville}},\ }\href@noop {} {\bibfield  {journal} {\bibinfo  {journal}
  {Commun. Math. Phys.}\ }\textbf {\bibinfo {volume} {74}},\ \bibinfo {pages}
  {189} (\bibinfo {year} {1980})}\BibitemShut {NoStop}%
\bibitem [{\citenamefont {Grebogi}\ \emph {et~al.}(1987)\citenamefont
  {Grebogi}, \citenamefont {Ott}, \citenamefont {Romeiras},\ and\ \citenamefont
  {Yorke}}]{Grebogi1987}%
  \BibitemOpen
  \bibfield  {author} {\bibinfo {author} {\bibfnamefont {C.}~\bibnamefont
  {Grebogi}}, \bibinfo {author} {\bibfnamefont {E.}~\bibnamefont {Ott}},
  \bibinfo {author} {\bibfnamefont {F.}~\bibnamefont {Romeiras}}, \ and\
  \bibinfo {author} {\bibfnamefont {J.~A.}\ \bibnamefont {Yorke}},\ }\href@noop
  {} {\bibfield  {journal} {\bibinfo  {journal} {Phys. Rev. A}\ }\textbf
  {\bibinfo {volume} {36}},\ \bibinfo {pages} {5365} (\bibinfo {year}
  {1987})}\BibitemShut {NoStop}%
\bibitem [{\citenamefont {Fujisaka}\ and\ \citenamefont
  {Yamada}(1985)}]{Fujisaka1985}%
  \BibitemOpen
  \bibfield  {author} {\bibinfo {author} {\bibfnamefont {H.}~\bibnamefont
  {Fujisaka}}\ and\ \bibinfo {author} {\bibfnamefont {T.}~\bibnamefont
  {Yamada}},\ }\href@noop {} {\bibfield  {journal} {\bibinfo  {journal} {Prog.
  Theor. Phys.}\ }\textbf {\bibinfo {volume} {74}},\ \bibinfo {pages} {918}
  (\bibinfo {year} {1985})}\BibitemShut {NoStop}%
\bibitem [{\citenamefont {Fujisaka}\ and\ \citenamefont
  {Yamada}(1986)}]{Fujisaka1986}%
  \BibitemOpen
  \bibfield  {author} {\bibinfo {author} {\bibfnamefont {H.}~\bibnamefont
  {Fujisaka}}\ and\ \bibinfo {author} {\bibfnamefont {T.}~\bibnamefont
  {Yamada}},\ }\href@noop {} {\bibfield  {journal} {\bibinfo  {journal} {Prog.
  Theor. Phys.}\ }\textbf {\bibinfo {volume} {75}},\ \bibinfo {pages} {1087}
  (\bibinfo {year} {1986})}\BibitemShut {NoStop}%
\bibitem [{\citenamefont {Platt}, \citenamefont {Spiegel},\ and\ \citenamefont
  {Tresser}(1993)}]{Platt1993}%
  \BibitemOpen
  \bibfield  {author} {\bibinfo {author} {\bibfnamefont {N.}~\bibnamefont
  {Platt}}, \bibinfo {author} {\bibfnamefont {E.~A.}\ \bibnamefont {Spiegel}},
  \ and\ \bibinfo {author} {\bibfnamefont {C.}~\bibnamefont {Tresser}},\
  }\href@noop {} {\bibfield  {journal} {\bibinfo  {journal} {Phys. Rev. Lett.}\
  }\textbf {\bibinfo {volume} {70}},\ \bibinfo {pages} {279} (\bibinfo {year}
  {1993})}\BibitemShut {NoStop}%
\bibitem [{\citenamefont {Heagy}, \citenamefont {Platt},\ and\ \citenamefont
  {Hammel}(1994)}]{Heagy1994}%
  \BibitemOpen
  \bibfield  {author} {\bibinfo {author} {\bibfnamefont {J.~F.}\ \bibnamefont
  {Heagy}}, \bibinfo {author} {\bibfnamefont {N.}~\bibnamefont {Platt}}, \ and\
  \bibinfo {author} {\bibfnamefont {S.~M.}\ \bibnamefont {Hammel}},\
  }\href@noop {} {\bibfield  {journal} {\bibinfo  {journal} {Phys. Rev. E}\
  }\textbf {\bibinfo {volume} {49}},\ \bibinfo {pages} {1140} (\bibinfo {year}
  {1994})}\BibitemShut {NoStop}%
\bibitem [{\citenamefont {Yamada}, \citenamefont {Fukushima},\ and\
  \citenamefont {Yazaki}(1989)}]{Yamada1989}%
  \BibitemOpen
  \bibfield  {author} {\bibinfo {author} {\bibfnamefont {T.}~\bibnamefont
  {Yamada}}, \bibinfo {author} {\bibfnamefont {K.}~\bibnamefont {Fukushima}}, \
  and\ \bibinfo {author} {\bibfnamefont {T.}~\bibnamefont {Yazaki}},\
  }\href@noop {} {\bibfield  {journal} {\bibinfo  {journal} {Prog. Theor. Phys.
  Suppl.}\ }\textbf {\bibinfo {volume} {99}},\ \bibinfo {pages} {120} (\bibinfo
  {year} {1989})}\BibitemShut {NoStop}%
\bibitem [{\citenamefont {Ott}\ and\ \citenamefont {Sommerer}(1994)}]{Ott1994}%
  \BibitemOpen
  \bibfield  {author} {\bibinfo {author} {\bibfnamefont {E.}~\bibnamefont
  {Ott}}\ and\ \bibinfo {author} {\bibfnamefont {J.~C.}\ \bibnamefont
  {Sommerer}},\ }\href@noop {} {\bibfield  {journal} {\bibinfo  {journal}
  {Phys. Lett. A}\ }\textbf {\bibinfo {volume} {188}},\ \bibinfo {pages} {39}
  (\bibinfo {year} {1994})}\BibitemShut {NoStop}%
\bibitem [{\citenamefont {Lai}\ and\ \citenamefont {Grebogi}(1995)}]{Lai1995}%
  \BibitemOpen
  \bibfield  {author} {\bibinfo {author} {\bibfnamefont {Y.-C.}\ \bibnamefont
  {Lai}}\ and\ \bibinfo {author} {\bibfnamefont {C.}~\bibnamefont {Grebogi}},\
  }\href@noop {} {\bibfield  {journal} {\bibinfo  {journal} {Phys. Rev. E}\
  }\textbf {\bibinfo {volume} {52}},\ \bibinfo {pages} {R3313} (\bibinfo {year}
  {1995})}\BibitemShut {NoStop}%
\bibitem [{\citenamefont {{\v{C}}enys}\ \emph {et~al.}(1996)\citenamefont
  {{\v{C}}enys}, \citenamefont {Namaj{\=u}nas}, \citenamefont
  {Tama{\v{s}}evi{\v{c}}ius},\ and\ \citenamefont {Schneider}}]{Cenys1996}%
  \BibitemOpen
  \bibfield  {author} {\bibinfo {author} {\bibfnamefont {A.}~\bibnamefont
  {{\v{C}}enys}}, \bibinfo {author} {\bibfnamefont {A.}~\bibnamefont
  {Namaj{\=u}nas}}, \bibinfo {author} {\bibfnamefont {A.}~\bibnamefont
  {Tama{\v{s}}evi{\v{c}}ius}}, \ and\ \bibinfo {author} {\bibfnamefont
  {T.}~\bibnamefont {Schneider}},\ }\href@noop {} {\bibfield  {journal}
  {\bibinfo  {journal} {Phys. Lett. A}\ }\textbf {\bibinfo {volume} {213}},\
  \bibinfo {pages} {259} (\bibinfo {year} {1996})}\BibitemShut {NoStop}%
\bibitem [{\citenamefont {Venkataramani}\ \emph {et~al.}(1996)\citenamefont
  {Venkataramani}, \citenamefont {T.~M.~Antonsen}, \citenamefont {Ott},\ and\
  \citenamefont {Sommerer}}]{Venkataramani1996}%
  \BibitemOpen
  \bibfield  {author} {\bibinfo {author} {\bibfnamefont {S.~C.}\ \bibnamefont
  {Venkataramani}}, \bibinfo {author} {\bibfnamefont {J.}~\bibnamefont
  {T.~M.~Antonsen}}, \bibinfo {author} {\bibfnamefont {E.}~\bibnamefont {Ott}},
  \ and\ \bibinfo {author} {\bibfnamefont {J.~C.}\ \bibnamefont {Sommerer}},\
  }\href@noop {} {\bibfield  {journal} {\bibinfo  {journal} {Physica D}\
  }\textbf {\bibinfo {volume} {96}},\ \bibinfo {pages} {66} (\bibinfo {year}
  {1996})}\BibitemShut {NoStop}%
\bibitem [{\citenamefont {Lai}(1996{\natexlab{a}})}]{Lai1996}%
  \BibitemOpen
  \bibfield  {author} {\bibinfo {author} {\bibfnamefont {Y.-C.}\ \bibnamefont
  {Lai}},\ }\href@noop {} {\bibfield  {journal} {\bibinfo  {journal} {Phys.
  Rev. E}\ }\textbf {\bibinfo {volume} {53}},\ \bibinfo {pages} {R4267}
  (\bibinfo {year} {1996}{\natexlab{a}})}\BibitemShut {NoStop}%
\bibitem [{\citenamefont {Lai}(1996{\natexlab{b}})}]{Lai1996a}%
  \BibitemOpen
  \bibfield  {author} {\bibinfo {author} {\bibfnamefont {Y.-C.}\ \bibnamefont
  {Lai}},\ }\href@noop {} {\bibfield  {journal} {\bibinfo  {journal} {Phys.
  Rev. E}\ }\textbf {\bibinfo {volume} {54}},\ \bibinfo {pages} {321} (\bibinfo
  {year} {1996}{\natexlab{b}})}\BibitemShut {NoStop}%
\bibitem [{\citenamefont {Fujisaka}\ \emph {et~al.}(1998)\citenamefont
  {Fujisaka}, \citenamefont {Ouchi}, \citenamefont {Hata}, \citenamefont
  {Masaoka},\ and\ \citenamefont {Miyazaki}}]{Fujisaka1998}%
  \BibitemOpen
  \bibfield  {author} {\bibinfo {author} {\bibfnamefont {H.}~\bibnamefont
  {Fujisaka}}, \bibinfo {author} {\bibfnamefont {K.}~\bibnamefont {Ouchi}},
  \bibinfo {author} {\bibfnamefont {H.}~\bibnamefont {Hata}}, \bibinfo {author}
  {\bibfnamefont {B.}~\bibnamefont {Masaoka}}, \ and\ \bibinfo {author}
  {\bibfnamefont {S.}~\bibnamefont {Miyazaki}},\ }\href@noop {} {\bibfield
  {journal} {\bibinfo  {journal} {Physica D}\ }\textbf {\bibinfo {volume}
  {114}},\ \bibinfo {pages} {237} (\bibinfo {year} {1998})}\BibitemShut
  {NoStop}%
\bibitem [{\citenamefont {Harada}, \citenamefont {Hata},\ and\ \citenamefont
  {Fujisaka}(1999)}]{Harada1999}%
  \BibitemOpen
  \bibfield  {author} {\bibinfo {author} {\bibfnamefont {T.}~\bibnamefont
  {Harada}}, \bibinfo {author} {\bibfnamefont {H.}~\bibnamefont {Hata}}, \ and\
  \bibinfo {author} {\bibfnamefont {H.}~\bibnamefont {Fujisaka}},\ }\href@noop
  {} {\bibfield  {journal} {\bibinfo  {journal} {J. Phys. A}\ }\textbf
  {\bibinfo {volume} {32}},\ \bibinfo {pages} {1557} (\bibinfo {year}
  {1999})}\BibitemShut {NoStop}%
\bibitem [{\citenamefont {Becker}\ \emph {et~al.}(1999)\citenamefont {Becker},
  \citenamefont {R{\"o}delsperger}, \citenamefont {Weyrauch}, \citenamefont
  {Benner}, \citenamefont {Just},\ and\ \citenamefont
  {{\v{C}}enys}}]{Becker1999}%
  \BibitemOpen
  \bibfield  {author} {\bibinfo {author} {\bibfnamefont {J.}~\bibnamefont
  {Becker}}, \bibinfo {author} {\bibfnamefont {F.}~\bibnamefont
  {R{\"o}delsperger}}, \bibinfo {author} {\bibfnamefont {T.}~\bibnamefont
  {Weyrauch}}, \bibinfo {author} {\bibfnamefont {H.}~\bibnamefont {Benner}},
  \bibinfo {author} {\bibfnamefont {W.}~\bibnamefont {Just}}, \ and\ \bibinfo
  {author} {\bibfnamefont {A.}~\bibnamefont {{\v{C}}enys}},\ }\href@noop {}
  {\bibfield  {journal} {\bibinfo  {journal} {Phys. Rev. E}\ }\textbf {\bibinfo
  {volume} {59}},\ \bibinfo {pages} {1622} (\bibinfo {year}
  {1999})}\BibitemShut {NoStop}%
\bibitem [{\citenamefont {Yamada}\ and\ \citenamefont
  {Fujisaka}(1986)}]{Yamada1986}%
  \BibitemOpen
  \bibfield  {author} {\bibinfo {author} {\bibfnamefont {T.}~\bibnamefont
  {Yamada}}\ and\ \bibinfo {author} {\bibfnamefont {H.}~\bibnamefont
  {Fujisaka}},\ }\href@noop {} {\bibfield  {journal} {\bibinfo  {journal}
  {Prog. Theor. Phys.}\ }\textbf {\bibinfo {volume} {76}},\ \bibinfo {pages}
  {582} (\bibinfo {year} {1986})}\BibitemShut {NoStop}%
\bibitem [{\citenamefont {Yamada}\ and\ \citenamefont
  {Fujisaka}(1990)}]{Yamada1990}%
  \BibitemOpen
  \bibfield  {author} {\bibinfo {author} {\bibfnamefont {T.}~\bibnamefont
  {Yamada}}\ and\ \bibinfo {author} {\bibfnamefont {H.}~\bibnamefont
  {Fujisaka}},\ }\href@noop {} {\bibfield  {journal} {\bibinfo  {journal}
  {Prog. Theor. Phys.}\ }\textbf {\bibinfo {volume} {84}},\ \bibinfo {pages}
  {824} (\bibinfo {year} {1990})}\BibitemShut {NoStop}%
\bibitem [{\citenamefont {Fujisaka}\ and\ \citenamefont
  {Yamada}(1987)}]{Fujisaka1987}%
  \BibitemOpen
  \bibfield  {author} {\bibinfo {author} {\bibfnamefont {H.}~\bibnamefont
  {Fujisaka}}\ and\ \bibinfo {author} {\bibfnamefont {T.}~\bibnamefont
  {Yamada}},\ }\href@noop {} {\bibfield  {journal} {\bibinfo  {journal} {Prog.
  Theor. Phys.}\ }\textbf {\bibinfo {volume} {77}},\ \bibinfo {pages} {1045}
  (\bibinfo {year} {1987})}\BibitemShut {NoStop}%
\bibitem [{\citenamefont {Fujisaka}\ and\ \citenamefont
  {Yamada}(1993)}]{Fujisaka1993}%
  \BibitemOpen
  \bibfield  {author} {\bibinfo {author} {\bibfnamefont {H.}~\bibnamefont
  {Fujisaka}}\ and\ \bibinfo {author} {\bibfnamefont {T.}~\bibnamefont
  {Yamada}},\ }\href@noop {} {\bibfield  {journal} {\bibinfo  {journal} {Prog.
  Theor. Phys.}\ }\textbf {\bibinfo {volume} {90}},\ \bibinfo {pages} {529}
  (\bibinfo {year} {1993})}\BibitemShut {NoStop}%
\bibitem [{\citenamefont {Suetani}\ and\ \citenamefont
  {Horita}(1999)}]{Suetani1999}%
  \BibitemOpen
  \bibfield  {author} {\bibinfo {author} {\bibfnamefont {H.}~\bibnamefont
  {Suetani}}\ and\ \bibinfo {author} {\bibfnamefont {T.}~\bibnamefont
  {Horita}},\ }\href@noop {} {\bibfield  {journal} {\bibinfo  {journal} {Phys.
  Rev. E}\ }\textbf {\bibinfo {volume} {60}},\ \bibinfo {pages} {422} (\bibinfo
  {year} {1999})}\BibitemShut {NoStop}%
\bibitem [{\citenamefont {Fujisaka}, \citenamefont {Suetani},\ and\
  \citenamefont {Watanabe}(2000)}]{Fujisaka2000}%
  \BibitemOpen
  \bibfield  {author} {\bibinfo {author} {\bibfnamefont {H.}~\bibnamefont
  {Fujisaka}}, \bibinfo {author} {\bibfnamefont {H.}~\bibnamefont {Suetani}}, \
  and\ \bibinfo {author} {\bibfnamefont {T.}~\bibnamefont {Watanabe}},\
  }\href@noop {} {\bibfield  {journal} {\bibinfo  {journal} {Prog. Theor. Phys.
  Suppl.}\ }\textbf {\bibinfo {volume} {139}},\ \bibinfo {pages} {70} (\bibinfo
  {year} {2000})}\BibitemShut {NoStop}%
\bibitem [{\citenamefont {Fujisaka}, \citenamefont {Matsushita},\ and\
  \citenamefont {Yamada}(1997)}]{Fujisaka1997}%
  \BibitemOpen
  \bibfield  {author} {\bibinfo {author} {\bibfnamefont {H.}~\bibnamefont
  {Fujisaka}}, \bibinfo {author} {\bibfnamefont {S.}~\bibnamefont
  {Matsushita}}, \ and\ \bibinfo {author} {\bibfnamefont {T.}~\bibnamefont
  {Yamada}},\ }\href@noop {} {\bibfield  {journal} {\bibinfo  {journal} {J.
  Phys. A}\ }\textbf {\bibinfo {volume} {30}},\ \bibinfo {pages} {5697}
  (\bibinfo {year} {1997})}\BibitemShut {NoStop}%
\bibitem [{\citenamefont {Miyazaki}(2000)}]{Miyazaki2000}%
  \BibitemOpen
  \bibfield  {author} {\bibinfo {author} {\bibfnamefont {S.}~\bibnamefont
  {Miyazaki}},\ }\href@noop {} {\bibfield  {journal} {\bibinfo  {journal} {J.
  Phys. Soc. Jpn.}\ }\textbf {\bibinfo {volume} {69}},\ \bibinfo {pages} {2719}
  (\bibinfo {year} {2000})}\BibitemShut {NoStop}%
\bibitem [{\citenamefont {Ward}\ and\ \citenamefont
  {Greenwood}(2007)}]{Scholarpedia2007}%
  \BibitemOpen
  \bibfield  {author} {\bibinfo {author} {\bibfnamefont {L.~M.}\ \bibnamefont
  {Ward}}\ and\ \bibinfo {author} {\bibfnamefont {P.~E.}\ \bibnamefont
  {Greenwood}},\ }\bibfield  {title} {\enquote {\bibinfo {title} {1/f noise},}\
  }\href@noop {} {\bibfield  {journal} {\bibinfo  {journal} {Scholarpedia}\
  }\textbf {\bibinfo {volume} {2}},\ \bibinfo {pages} {1537} (\bibinfo {year}
  {2007})}\BibitemShut {NoStop}%
\bibitem [{\citenamefont {Weissman}(1988)}]{Weissman1988}%
  \BibitemOpen
  \bibfield  {author} {\bibinfo {author} {\bibfnamefont {M.~B.}\ \bibnamefont
  {Weissman}},\ }\href@noop {} {\bibfield  {journal} {\bibinfo  {journal} {Rev.
  Mod. Phys.}\ }\textbf {\bibinfo {volume} {60}},\ \bibinfo {pages} {537}
  (\bibinfo {year} {1988})}\BibitemShut {NoStop}%
\bibitem [{\citenamefont {Barabasi}\ and\ \citenamefont
  {Albert}(1999)}]{Barabasi1999}%
  \BibitemOpen
  \bibfield  {author} {\bibinfo {author} {\bibfnamefont {A.~L.}\ \bibnamefont
  {Barabasi}}\ and\ \bibinfo {author} {\bibfnamefont {R.}~\bibnamefont
  {Albert}},\ }\href@noop {} {\bibfield  {journal} {\bibinfo  {journal}
  {Science}\ }\textbf {\bibinfo {volume} {286}},\ \bibinfo {pages} {509}
  (\bibinfo {year} {1999})}\BibitemShut {NoStop}%
\bibitem [{\citenamefont {Gisiger}(2001)}]{Gisiger2001}%
  \BibitemOpen
  \bibfield  {author} {\bibinfo {author} {\bibfnamefont {T.}~\bibnamefont
  {Gisiger}},\ }\href@noop {} {\bibfield  {journal} {\bibinfo  {journal} {Biol.
  Rev.}\ }\textbf {\bibinfo {volume} {76}},\ \bibinfo {pages} {161} (\bibinfo
  {year} {2001})}\BibitemShut {NoStop}%
\bibitem [{\citenamefont {Wong}(2003)}]{Wong2003}%
  \BibitemOpen
  \bibfield  {author} {\bibinfo {author} {\bibfnamefont {H.}~\bibnamefont
  {Wong}},\ }\href@noop {} {\bibfield  {journal} {\bibinfo  {journal}
  {Microelectron. Reliab.}\ }\textbf {\bibinfo {volume} {43}},\ \bibinfo
  {pages} {585} (\bibinfo {year} {2003})}\BibitemShut {NoStop}%
\bibitem [{\citenamefont {Wagenmakers}, \citenamefont {Farrell},\ and\
  \citenamefont {Ratcliff}(2004)}]{Wagenmakers2004}%
  \BibitemOpen
  \bibfield  {author} {\bibinfo {author} {\bibfnamefont {E.-J.}\ \bibnamefont
  {Wagenmakers}}, \bibinfo {author} {\bibfnamefont {S.}~\bibnamefont
  {Farrell}}, \ and\ \bibinfo {author} {\bibfnamefont {R.}~\bibnamefont
  {Ratcliff}},\ }\href@noop {} {\bibfield  {journal} {\bibinfo  {journal}
  {Psychonomic Bull. Rev.}\ }\textbf {\bibinfo {volume} {11}},\ \bibinfo
  {pages} {579} (\bibinfo {year} {2004})}\BibitemShut {NoStop}%
\bibitem [{\citenamefont {Newman}(2005)}]{Newman05}%
  \BibitemOpen
  \bibfield  {author} {\bibinfo {author} {\bibfnamefont {M.~E.~J.}\
  \bibnamefont {Newman}},\ }\href@noop {} {\bibfield  {journal} {\bibinfo
  {journal} {Contemp. Phys.}\ }\textbf {\bibinfo {volume} {46}},\ \bibinfo
  {pages} {323} (\bibinfo {year} {2005})}\BibitemShut {NoStop}%
\bibitem [{\citenamefont {Szabo}\ and\ \citenamefont {Fath}(2007)}]{Szabo2007}%
  \BibitemOpen
  \bibfield  {author} {\bibinfo {author} {\bibfnamefont {G.}~\bibnamefont
  {Szabo}}\ and\ \bibinfo {author} {\bibfnamefont {G.}~\bibnamefont {Fath}},\
  }\href@noop {} {\bibfield  {journal} {\bibinfo  {journal} {Phys. Rep.}\
  }\textbf {\bibinfo {volume} {446}},\ \bibinfo {pages} {97} (\bibinfo {year}
  {2007})}\BibitemShut {NoStop}%
\bibitem [{\citenamefont {Castellano}, \citenamefont {Fortunato},\ and\
  \citenamefont {Loreto}(2009)}]{Castellano2009}%
  \BibitemOpen
  \bibfield  {author} {\bibinfo {author} {\bibfnamefont {C.}~\bibnamefont
  {Castellano}}, \bibinfo {author} {\bibfnamefont {S.}~\bibnamefont
  {Fortunato}}, \ and\ \bibinfo {author} {\bibfnamefont {V.}~\bibnamefont
  {Loreto}},\ }\href@noop {} {\bibfield  {journal} {\bibinfo  {journal} {Rev.
  Mod. Phys.}\ }\textbf {\bibinfo {volume} {81}},\ \bibinfo {pages} {591}
  (\bibinfo {year} {2009})}\BibitemShut {NoStop}%
\bibitem [{\citenamefont {Eliazar}\ and\ \citenamefont
  {Klafter}(2009)}]{Eliazar2009}%
  \BibitemOpen
  \bibfield  {author} {\bibinfo {author} {\bibfnamefont {I.~E.}\ \bibnamefont
  {Eliazar}}\ and\ \bibinfo {author} {\bibfnamefont {J.}~\bibnamefont
  {Klafter}},\ }\href@noop {} {\bibfield  {journal} {\bibinfo  {journal} {Proc.
  Natl. Acad. Sci. U.S.A.}\ }\textbf {\bibinfo {volume} {106}},\ \bibinfo
  {pages} {12251} (\bibinfo {year} {2009})}\BibitemShut {NoStop}%
\bibitem [{\citenamefont {Eliazar}\ and\ \citenamefont
  {Klafter}(2010)}]{Eliazar2010}%
  \BibitemOpen
  \bibfield  {author} {\bibinfo {author} {\bibfnamefont {I.~E.}\ \bibnamefont
  {Eliazar}}\ and\ \bibinfo {author} {\bibfnamefont {J.}~\bibnamefont
  {Klafter}},\ }\href@noop {} {\bibfield  {journal} {\bibinfo  {journal} {Phys.
  Rev. E}\ }\textbf {\bibinfo {volume} {82}},\ \bibinfo {pages} {021109}
  (\bibinfo {year} {2010})}\BibitemShut {NoStop}%
\bibitem [{\citenamefont {Perc}\ and\ \citenamefont
  {Szolnoki}(2010)}]{Perc2010}%
  \BibitemOpen
  \bibfield  {author} {\bibinfo {author} {\bibfnamefont {M.}~\bibnamefont
  {Perc}}\ and\ \bibinfo {author} {\bibfnamefont {A.}~\bibnamefont
  {Szolnoki}},\ }\href@noop {} {\bibfield  {journal} {\bibinfo  {journal}
  {Biosystems}\ }\textbf {\bibinfo {volume} {99}},\ \bibinfo {pages} {109}
  (\bibinfo {year} {2010})}\BibitemShut {NoStop}%
\bibitem [{\citenamefont {Werner}(2010)}]{Werner2010}%
  \BibitemOpen
  \bibfield  {author} {\bibinfo {author} {\bibfnamefont {G.}~\bibnamefont
  {Werner}},\ }\href@noop {} {\bibfield  {journal} {\bibinfo  {journal}
  {Frontiers in Physiology}\ }\textbf {\bibinfo {volume} {1}},\ \bibinfo
  {pages} {1} (\bibinfo {year} {2010})}\BibitemShut {NoStop}%
\bibitem [{\citenamefont {Orden}(2010)}]{Orden2010}%
  \BibitemOpen
  \bibfield  {author} {\bibinfo {author} {\bibfnamefont {G.~V.}\ \bibnamefont
  {Orden}},\ }\href@noop {} {\bibfield  {journal} {\bibinfo  {journal}
  {Medicina (Kaunas)}\ }\textbf {\bibinfo {volume} {46}},\ \bibinfo {pages}
  {581} (\bibinfo {year} {2010})}\BibitemShut {NoStop}%
\bibitem [{\citenamefont {Kendal}\ and\ \citenamefont
  {Jorgensen}(2011)}]{Kendal2011}%
  \BibitemOpen
  \bibfield  {author} {\bibinfo {author} {\bibfnamefont {W.~S.}\ \bibnamefont
  {Kendal}}\ and\ \bibinfo {author} {\bibfnamefont {B.}~\bibnamefont
  {Jorgensen}},\ }\href@noop {} {\bibfield  {journal} {\bibinfo  {journal}
  {Phys. Rev. E}\ }\textbf {\bibinfo {volume} {84}},\ \bibinfo {pages} {066120}
  (\bibinfo {year} {2011})}\BibitemShut {NoStop}%
\bibitem [{\citenamefont {Torabi}\ and\ \citenamefont
  {Berg}(2011)}]{Torabi2011}%
  \BibitemOpen
  \bibfield  {author} {\bibinfo {author} {\bibfnamefont {A.}~\bibnamefont
  {Torabi}}\ and\ \bibinfo {author} {\bibfnamefont {S.~S.}\ \bibnamefont
  {Berg}},\ }\href@noop {} {\bibfield  {journal} {\bibinfo  {journal} {Marine
  and Petroleum Geology}\ }\textbf {\bibinfo {volume} {28}},\ \bibinfo {pages}
  {1444} (\bibinfo {year} {2011})}\BibitemShut {NoStop}%
\bibitem [{\citenamefont {Diniz}\ \emph {et~al.}(2011)\citenamefont {Diniz},
  \citenamefont {Wijnants}, \citenamefont {Torre}, \citenamefont {Barreiros},
  \citenamefont {Crato}, \citenamefont {Bosman}, \citenamefont {Hasselman},
  \citenamefont {Cox}, \citenamefont {Orden},\ and\ \citenamefont
  {Deligni�res}}]{Diniz2011}%
  \BibitemOpen
  \bibfield  {author} {\bibinfo {author} {\bibfnamefont {A.}~\bibnamefont
  {Diniz}}, \bibinfo {author} {\bibfnamefont {M.~L.}\ \bibnamefont {Wijnants}},
  \bibinfo {author} {\bibfnamefont {K.}~\bibnamefont {Torre}}, \bibinfo
  {author} {\bibfnamefont {J.}~\bibnamefont {Barreiros}}, \bibinfo {author}
  {\bibfnamefont {N.}~\bibnamefont {Crato}}, \bibinfo {author} {\bibfnamefont
  {A.~M.~T.}\ \bibnamefont {Bosman}}, \bibinfo {author} {\bibfnamefont
  {F.}~\bibnamefont {Hasselman}}, \bibinfo {author} {\bibfnamefont {R.~F.}\
  \bibnamefont {Cox}}, \bibinfo {author} {\bibfnamefont {G.~C.~V.}\
  \bibnamefont {Orden}}, \ and\ \bibinfo {author} {\bibfnamefont
  {D.}~\bibnamefont {Deligni�res}},\ }\href@noop {} {\bibfield  {journal}
  {\bibinfo  {journal} {Human Movement Science}\ }\textbf {\bibinfo {volume}
  {30}},\ \bibinfo {pages} {889} (\bibinfo {year} {2011})}\BibitemShut
  {NoStop}%
\bibitem [{\citenamefont {Kaulakys}\ and\ \citenamefont
  {Ruseckas}(2004)}]{Kaulakys2004}%
  \BibitemOpen
  \bibfield  {author} {\bibinfo {author} {\bibfnamefont {B.}~\bibnamefont
  {Kaulakys}}\ and\ \bibinfo {author} {\bibfnamefont {J.}~\bibnamefont
  {Ruseckas}},\ }\href@noop {} {\bibfield  {journal} {\bibinfo  {journal}
  {Phys. Rev. E}\ }\textbf {\bibinfo {volume} {70}},\ \bibinfo {pages}
  {020101(R)} (\bibinfo {year} {2004})}\BibitemShut {NoStop}%
\bibitem [{\citenamefont {Kaulakys}\ \emph {et~al.}(2006)\citenamefont
  {Kaulakys}, \citenamefont {Ruseckas}, \citenamefont {Gontis},\ and\
  \citenamefont {Alaburda}}]{Kaulakys2006}%
  \BibitemOpen
  \bibfield  {author} {\bibinfo {author} {\bibfnamefont {B.}~\bibnamefont
  {Kaulakys}}, \bibinfo {author} {\bibfnamefont {J.}~\bibnamefont {Ruseckas}},
  \bibinfo {author} {\bibfnamefont {V.}~\bibnamefont {Gontis}}, \ and\ \bibinfo
  {author} {\bibfnamefont {M.}~\bibnamefont {Alaburda}},\ }\href@noop {}
  {\bibfield  {journal} {\bibinfo  {journal} {Physica A}\ }\textbf {\bibinfo
  {volume} {365}},\ \bibinfo {pages} {217} (\bibinfo {year}
  {2006})}\BibitemShut {NoStop}%
\bibitem [{\citenamefont {Kaulakys}\ and\ \citenamefont
  {Alaburda}(2009)}]{Kaulakys2009}%
  \BibitemOpen
  \bibfield  {author} {\bibinfo {author} {\bibfnamefont {B.}~\bibnamefont
  {Kaulakys}}\ and\ \bibinfo {author} {\bibfnamefont {M.}~\bibnamefont
  {Alaburda}},\ }\href@noop {} {\bibfield  {journal} {\bibinfo  {journal} {J.
  Stat. Mech.}\ }\textbf {\bibinfo {volume} {2009}},\ \bibinfo {pages} {P02051}
  (\bibinfo {year} {2009})}\BibitemShut {NoStop}%
\bibitem [{\citenamefont {Ruseckas}\ and\ \citenamefont
  {Kaulakys}(2010)}]{Ruseckas10}%
  \BibitemOpen
  \bibfield  {author} {\bibinfo {author} {\bibfnamefont {J.}~\bibnamefont
  {Ruseckas}}\ and\ \bibinfo {author} {\bibfnamefont {B.}~\bibnamefont
  {Kaulakys}},\ }\href@noop {} {\bibfield  {journal} {\bibinfo  {journal}
  {Phys. Rev. E}\ }\textbf {\bibinfo {volume} {81}},\ \bibinfo {pages} {031105}
  (\bibinfo {year} {2010})}\BibitemShut {NoStop}%
\bibitem [{\citenamefont {Kaulakys}\ and\ \citenamefont
  {Me\v{s}kauskas}(1998)}]{Kaulakys1998-1}%
  \BibitemOpen
  \bibfield  {author} {\bibinfo {author} {\bibfnamefont {B.}~\bibnamefont
  {Kaulakys}}\ and\ \bibinfo {author} {\bibfnamefont {T.}~\bibnamefont
  {Me\v{s}kauskas}},\ }\href@noop {} {\bibfield  {journal} {\bibinfo  {journal}
  {Phys. Rev. E}\ }\textbf {\bibinfo {volume} {58}},\ \bibinfo {pages} {7013}
  (\bibinfo {year} {1998})}\BibitemShut {NoStop}%
\bibitem [{\citenamefont {Kaulakys}(1999)}]{Kaulakys1999-1}%
  \BibitemOpen
  \bibfield  {author} {\bibinfo {author} {\bibfnamefont {B.}~\bibnamefont
  {Kaulakys}},\ }\href@noop {} {\bibfield  {journal} {\bibinfo  {journal}
  {Phys. Lett. A}\ }\textbf {\bibinfo {volume} {257}},\ \bibinfo {pages} {37}
  (\bibinfo {year} {1999})}\BibitemShut {NoStop}%
\bibitem [{\citenamefont {Kaulakys}\ and\ \citenamefont
  {Me\v{s}kauskas}(2000)}]{Kaulakys2000-1}%
  \BibitemOpen
  \bibfield  {author} {\bibinfo {author} {\bibfnamefont {B.}~\bibnamefont
  {Kaulakys}}\ and\ \bibinfo {author} {\bibfnamefont {T.}~\bibnamefont
  {Me\v{s}kauskas}},\ }\href@noop {} {\bibfield  {journal} {\bibinfo  {journal}
  {Microel. Reliab.}\ }\textbf {\bibinfo {volume} {40}},\ \bibinfo {pages}
  {1781} (\bibinfo {year} {2000})}\BibitemShut {NoStop}%
\bibitem [{\citenamefont {Kaulakys}(2000)}]{Kaulakys2000-2}%
  \BibitemOpen
  \bibfield  {author} {\bibinfo {author} {\bibfnamefont {B.}~\bibnamefont
  {Kaulakys}},\ }\href@noop {} {\bibfield  {journal} {\bibinfo  {journal}
  {Microel. Reliab.}\ }\textbf {\bibinfo {volume} {40}},\ \bibinfo {pages}
  {1787} (\bibinfo {year} {2000})}\BibitemShut {NoStop}%
\bibitem [{\citenamefont {Gontis}\ and\ \citenamefont
  {Kaulakys}(2004)}]{Gontis2004}%
  \BibitemOpen
  \bibfield  {author} {\bibinfo {author} {\bibfnamefont {V.}~\bibnamefont
  {Gontis}}\ and\ \bibinfo {author} {\bibfnamefont {B.}~\bibnamefont
  {Kaulakys}},\ }\href@noop {} {\bibfield  {journal} {\bibinfo  {journal}
  {Physica A}\ }\textbf {\bibinfo {volume} {343}},\ \bibinfo {pages} {505}
  (\bibinfo {year} {2004})}\BibitemShut {NoStop}%
\bibitem [{\citenamefont {Kaulakys}, \citenamefont {Gontis},\ and\
  \citenamefont {Alaburda}(2005)}]{Kaulakys2005}%
  \BibitemOpen
  \bibfield  {author} {\bibinfo {author} {\bibfnamefont {B.}~\bibnamefont
  {Kaulakys}}, \bibinfo {author} {\bibfnamefont {V.}~\bibnamefont {Gontis}}, \
  and\ \bibinfo {author} {\bibfnamefont {M.}~\bibnamefont {Alaburda}},\
  }\href@noop {} {\bibfield  {journal} {\bibinfo  {journal} {Phys. Rev. E}\
  }\textbf {\bibinfo {volume} {71}},\ \bibinfo {pages} {051105} (\bibinfo
  {year} {2005})}\BibitemShut {NoStop}%
\bibitem [{\citenamefont {Procaccia}\ and\ \citenamefont
  {Schuster}(1983)}]{Procaccia1983}%
  \BibitemOpen
  \bibfield  {author} {\bibinfo {author} {\bibfnamefont {I.}~\bibnamefont
  {Procaccia}}\ and\ \bibinfo {author} {\bibfnamefont {H.}~\bibnamefont
  {Schuster}},\ }\href@noop {} {\bibfield  {journal} {\bibinfo  {journal}
  {Phys. Rev. A}\ }\textbf {\bibinfo {volume} {28}},\ \bibinfo {pages} {1210}
  (\bibinfo {year} {1983})}\BibitemShut {NoStop}%
\bibitem [{\citenamefont {Schuster}(1988)}]{Schuster1988}%
  \BibitemOpen
  \bibfield  {author} {\bibinfo {author} {\bibfnamefont {H.~G.}\ \bibnamefont
  {Schuster}},\ }\href@noop {} {\emph {\bibinfo {title} {Deterministic
  Chaos}}}\ (\bibinfo  {publisher} {VCH},\ \bibinfo {address} {Weinheim},\
  \bibinfo {year} {1988})\BibitemShut {NoStop}%
\bibitem [{\citenamefont {Costa}\ \emph {et~al.}(1997)\citenamefont {Costa},
  \citenamefont {Lyra}, \citenamefont {Plastino},\ and\ \citenamefont
  {Tsallis}}]{Costa1997}%
  \BibitemOpen
  \bibfield  {author} {\bibinfo {author} {\bibfnamefont {U.~M.~S.}\
  \bibnamefont {Costa}}, \bibinfo {author} {\bibfnamefont {M.~L.}\ \bibnamefont
  {Lyra}}, \bibinfo {author} {\bibfnamefont {A.~R.}\ \bibnamefont {Plastino}},
  \ and\ \bibinfo {author} {\bibfnamefont {C.}~\bibnamefont {Tsallis}},\
  }\href@noop {} {\bibfield  {journal} {\bibinfo  {journal} {Phys. Rev. E}\
  }\textbf {\bibinfo {volume} {56}},\ \bibinfo {pages} {245} (\bibinfo {year}
  {1997})}\BibitemShut {NoStop}%
\bibitem [{\citenamefont {Manneville}(1980)}]{Manneville1980a}%
  \BibitemOpen
  \bibfield  {author} {\bibinfo {author} {\bibfnamefont {P.}~\bibnamefont
  {Manneville}},\ }\href@noop {} {\bibfield  {journal} {\bibinfo  {journal} {J.
  Physique (Paris)}\ }\textbf {\bibinfo {volume} {41}},\ \bibinfo {pages}
  {1235} (\bibinfo {year} {1980})}\BibitemShut {NoStop}%
\bibitem [{\citenamefont {Ben-Mizrachi}\ \emph {et~al.}(1985)\citenamefont
  {Ben-Mizrachi}, \citenamefont {Procaccia}, \citenamefont {Rosenberg},
  \citenamefont {Schmidt},\ and\ \citenamefont {Schuster}}]{Ben-Mizrachi1985}%
  \BibitemOpen
  \bibfield  {author} {\bibinfo {author} {\bibfnamefont {A.}~\bibnamefont
  {Ben-Mizrachi}}, \bibinfo {author} {\bibfnamefont {I.}~\bibnamefont
  {Procaccia}}, \bibinfo {author} {\bibfnamefont {N.}~\bibnamefont
  {Rosenberg}}, \bibinfo {author} {\bibfnamefont {A.}~\bibnamefont {Schmidt}},
  \ and\ \bibinfo {author} {\bibfnamefont {H.~G.}\ \bibnamefont {Schuster}},\
  }\href@noop {} {\bibfield  {journal} {\bibinfo  {journal} {Phys. Rev. A}\
  }\textbf {\bibinfo {volume} {31}},\ \bibinfo {pages} {1830} (\bibinfo {year}
  {1985})}\BibitemShut {NoStop}%
\bibitem [{\citenamefont {Laurson}\ and\ \citenamefont
  {Alava}(2006)}]{Laurson2006}%
  \BibitemOpen
  \bibfield  {author} {\bibinfo {author} {\bibfnamefont {L.}~\bibnamefont
  {Laurson}}\ and\ \bibinfo {author} {\bibfnamefont {M.~J.}\ \bibnamefont
  {Alava}},\ }\href@noop {} {\bibfield  {journal} {\bibinfo  {journal} {Phys.
  Rev. E}\ }\textbf {\bibinfo {volume} {74}},\ \bibinfo {pages} {066106}
  (\bibinfo {year} {2006})}\BibitemShut {NoStop}%
\bibitem [{\citenamefont {Pando~L.}\ and\ \citenamefont
  {Doedel}(2007)}]{Pando2007}%
  \BibitemOpen
  \bibfield  {author} {\bibinfo {author} {\bibfnamefont {C.~L.}\ \bibnamefont
  {Pando~L.}}\ and\ \bibinfo {author} {\bibfnamefont {E.~J.}\ \bibnamefont
  {Doedel}},\ }\href@noop {} {\bibfield  {journal} {\bibinfo  {journal} {Phys.
  Rev. E}\ }\textbf {\bibinfo {volume} {75}},\ \bibinfo {pages} {016213}
  (\bibinfo {year} {2007})}\BibitemShut {NoStop}%
\bibitem [{\citenamefont {Shinkai}\ and\ \citenamefont
  {Aizawa}(2012)}]{Shinkai2012}%
  \BibitemOpen
  \bibfield  {author} {\bibinfo {author} {\bibfnamefont {S.}~\bibnamefont
  {Shinkai}}\ and\ \bibinfo {author} {\bibfnamefont {Y.}~\bibnamefont
  {Aizawa}},\ }\href@noop {} {\bibfield  {journal} {\bibinfo  {journal} {J.
  Phys. Soc. Jpn.}\ }\textbf {\bibinfo {volume} {81}},\ \bibinfo {pages}
  {024009} (\bibinfo {year} {2012})}\BibitemShut {NoStop}%
\bibitem [{\citenamefont {Zaslavsky}(2007)}]{Zaslavsky2007}%
  \BibitemOpen
  \bibfield  {author} {\bibinfo {author} {\bibfnamefont {G.~M.}\ \bibnamefont
  {Zaslavsky}},\ }\href@noop {} {\emph {\bibinfo {title} {The Physics of Chaos
  in Hamiltonian Systems}}},\ \bibinfo {edition} {2nd}\ ed.\ (\bibinfo
  {publisher} {Imperial College Pres},\ \bibinfo {address} {London},\ \bibinfo
  {year} {2007})\BibitemShut {NoStop}%
\bibitem [{\citenamefont {Tsallis}(1988)}]{Tsallis1988}%
  \BibitemOpen
  \bibfield  {author} {\bibinfo {author} {\bibfnamefont {C.}~\bibnamefont
  {Tsallis}},\ }\href@noop {} {\bibfield  {journal} {\bibinfo  {journal} {J.
  Stat. Phys.}\ }\textbf {\bibinfo {volume} {52}},\ \bibinfo {pages} {479}
  (\bibinfo {year} {1988})}\BibitemShut {NoStop}%
\bibitem [{\citenamefont {Tsallis}(2009{\natexlab{a}})}]{Tsallis2009}%
  \BibitemOpen
  \bibfield  {author} {\bibinfo {author} {\bibfnamefont {C.}~\bibnamefont
  {Tsallis}},\ }\href@noop {} {\emph {\bibinfo {title} {Introduction to
  Nonextensive Statistical Mechanics: Approaching a Complex World}}}\ (\bibinfo
   {publisher} {Springer},\ \bibinfo {address} {New York},\ \bibinfo {year}
  {2009})\BibitemShut {NoStop}%
\bibitem [{\citenamefont {Tsallis}(2009{\natexlab{b}})}]{Tsallis2009a}%
  \BibitemOpen
  \bibfield  {author} {\bibinfo {author} {\bibfnamefont {C.}~\bibnamefont
  {Tsallis}},\ }\href@noop {} {\bibfield  {journal} {\bibinfo  {journal} {Braz.
  J. Phys.}\ }\textbf {\bibinfo {volume} {39}},\ \bibinfo {pages} {337}
  (\bibinfo {year} {2009}{\natexlab{b}})}\BibitemShut {NoStop}%
\bibitem [{\citenamefont {Redner}(2001)}]{Redner2001}%
  \BibitemOpen
  \bibfield  {author} {\bibinfo {author} {\bibfnamefont {S.}~\bibnamefont
  {Redner}},\ }\href@noop {} {\emph {\bibinfo {title} {A Guide to First-Passage
  Processes}}}\ (\bibinfo  {publisher} {Cambridge University Press},\ \bibinfo
  {year} {2001})\BibitemShut {NoStop}%
\bibitem [{\citenamefont {Ruseckas}, \citenamefont {Kaulakys},\ and\
  \citenamefont {Gontis}(2011)}]{ruseckas-2011epl}%
  \BibitemOpen
  \bibfield  {author} {\bibinfo {author} {\bibfnamefont {J.}~\bibnamefont
  {Ruseckas}}, \bibinfo {author} {\bibfnamefont {B.}~\bibnamefont {Kaulakys}},
  \ and\ \bibinfo {author} {\bibfnamefont {V.}~\bibnamefont {Gontis}},\
  }\href@noop {} {\bibfield  {journal} {\bibinfo  {journal} {EPL}\ }\textbf
  {\bibinfo {volume} {96}},\ \bibinfo {pages} {60007} (\bibinfo {year}
  {2011})}\BibitemShut {NoStop}%
\bibitem [{\citenamefont {Ruseckas}\ and\ \citenamefont
  {Kaulakys}(2011)}]{ruseckas-2011}%
  \BibitemOpen
  \bibfield  {author} {\bibinfo {author} {\bibfnamefont {J.}~\bibnamefont
  {Ruseckas}}\ and\ \bibinfo {author} {\bibfnamefont {B.}~\bibnamefont
  {Kaulakys}},\ }\href@noop {} {\bibfield  {journal} {\bibinfo  {journal}
  {Phys. Rev. E}\ }\textbf {\bibinfo {volume} {84}},\ \bibinfo {pages} {051125}
  (\bibinfo {year} {2011})}\BibitemShut {NoStop}%
\bibitem [{\citenamefont {Risken}\ and\ \citenamefont
  {Frank}(1996)}]{Risken1996}%
  \BibitemOpen
  \bibfield  {author} {\bibinfo {author} {\bibfnamefont {H.}~\bibnamefont
  {Risken}}\ and\ \bibinfo {author} {\bibfnamefont {T.}~\bibnamefont {Frank}},\
  }\href@noop {} {\emph {\bibinfo {title} {The Fokker-Planck Equation: Methods
  of Solution and Applications}}}\ (\bibinfo  {publisher} {Springer},\ \bibinfo
  {year} {1996})\BibitemShut {NoStop}%
\bibitem [{\citenamefont {Gardiner}(2004)}]{Gardiner04}%
  \BibitemOpen
  \bibfield  {author} {\bibinfo {author} {\bibfnamefont {C.~W.}\ \bibnamefont
  {Gardiner}},\ }\href@noop {} {\emph {\bibinfo {title} {Handbook of Stochastic
  Methods for Physics, Chemistry and the Natural Sciences}}}\ (\bibinfo
  {publisher} {Springer-Verlag},\ \bibinfo {address} {Berlin},\ \bibinfo {year}
  {2004})\BibitemShut {NoStop}%
\end{thebibliography}

%

\end{document}